\documentclass[12pt, draftclsnofoot, onecolumn]{IEEEtran}
\ifCLASSINFOpdf

\else
\fi
\usepackage{url}
\usepackage{comment}
\includecomment{toexclude}
\usepackage{amsmath,amsfonts,amssymb}
\usepackage{cite}
\usepackage{verbatim}
\usepackage{bm}
\usepackage{graphicx}
\usepackage{xcolor}
\usepackage{subfigure}
\usepackage{diagbox}
\usepackage{multirow}
\usepackage{array}
\usepackage{statmath}
\usepackage[belowskip=-5pt,aboveskip=0pt]{caption}
\newcolumntype{P}[1]{>{\centering\arraybackslash}p{#1}}
\usepackage{flushend}

\usepackage{enumitem}
\usepackage{epstopdf}
\usepackage{algorithm,algpseudocode}
\usepackage[justification=centering]{caption}

\usepackage{xcolor}
\graphicspath{{figures/}}

\newcommand{\Gk}{{\bf \hat{G}}_k}

\usepackage{mathtools}
\DeclarePairedDelimiter\floor{\lfloor}{\rfloor}

\algrenewcommand\algorithmicindent{0.7em}%
\DeclareMathAlphabet\mathbfcal{OMS}{cmsy}{b}{n}
\usepackage{stackengine}
\def\delequal{\mathrel{\ensurestackMath{\stackon[1pt]{=}{\scriptstyle\Delta}}}}
\begin{document}

\title{Multiple Correlated Jammers Nullification using LSTM-based Deep Dueling Neural Network}

\author{Linh~Manh~Hoang,~\IEEEmembership{Student Member,~IEEE}, Diep~N.~Nguyen,~\IEEEmembership{Senior Member,~IEEE},
	J.~Andrew~Zhang,~\IEEEmembership{Senior Member,~IEEE}, and Dinh~Thai~Hoang,~\IEEEmembership{Member,~IEEE}
	\thanks{ L. M. Hoang, D. N. Nguyen, J. A. Zhang and D. T. Hoang are with the School of Electrical and Data Engineering, University of Technology Sydney, NSW 2007 (e-mail: Linh.M.Hoang@student.uts.edu.au; Diep.Nguyen@uts.edu.au; Andrew.Zhang@uts.edu.au; Hoang.Dinh@uts.edu.au).
		
	Preliminary results of this work will be presented at the IEEE WCNC Conference 2022\cite{hoang2021multiple}.}
}
{}
\maketitle
\vspace{-1cm}
\begin{abstract}
Suppressing the inadvertent or deliberate interference for wireless networks is critical to guarantee a reliable communication link. However, nullifying the jamming signals can be problematic when the correlations between transmitted jamming signals are deliberately varied over time. Specifically, recent studies reveal that by deliberately varying the correlations among jamming signals, attackers can effectively vary the jamming channels and thus their nullspace, even when the physical channels remain unchanged. That makes the estimated beam-forming matrix derived from the nullspace of the jamming channels unable to suppress the jamming signals. Most existing interference nullification solutions only consider unchanged correlations or heuristically adapt to the time-varying correlation problem by continuously monitoring the residual jamming signals before updating the estimated beam-forming matrix. In this paper, we systematically formulate the optimization problem of the nullspace estimation and data transmission phases. Even ignoring the unknown strategy of the jammers and the challenging nullspace estimation process, the resulting problem is an integer programming problem, hence intractable to obtain its optimal solution. To tackle it and address the unknown strategy of the jammer, we reformulate the problem using a partially observable semi-Markov decision process (POSMDP) and then design a deep dueling Q-learning based framework to tune the duration of the nullspace estimation and data transmission phases. As such, the estimated beam-forming matrix remains effective even when the correlations are time-varying. Extensive simulations demonstrate that the proposed techniques effectively deal with jamming signals whose correlations vary over time, and the range of correlations is unknown. Especially, our techniques do not require continuous monitoring of the residual jamming signals (after the nullification process) before updating the estimated beam-forming matrix. As such, the system is more spectral-efficient and has a lower outage probability.
\end{abstract}

%
\IEEEpeerreviewmaketitle
\vspace{-0.5cm}
\section{Introduction}
The angle of arrival (AoA)-based beam-forming is among the most popular techniques to suppress jamming signals. It is accomplished by estimating the AoAs of the incoming spatial streams of the jamming signals using AoA estimation techniques, such as the multiple signal classification (MUSIC)\cite{schmidt1986multiple}, matrix pencil \cite{hua1990matrix}, and estimation of signal parameters via rotational invariance techniques (ESPRIT)\cite{roy1989esprit}. These AoA estimation techniques can be combined with the spatial smoothing technique \cite{chen2010introduction} to deal with the correlated jamming signals. However, for this technique, at least one degree-of-freedom is needed to nullify each propagation path of the jamming signals\cite{fenn2007adaptive}. Hence, this approach is only applicable when the number of jammers is small, and the environment is sparse multipath. In the rich scattering environment where the number of propagation paths is large, the receiver has to ``sacrifice'' many degree-of-freedoms for jamming suppression, resulting in a significantly lower throughput\cite{lu2014overview}.


Spread spectrum communication is another approach to deal with jamming signals. Specifically, in the frequency-hopping spread spectrum (FHSS) technique \cite{popper2010anti,7524449,gao2018game,wang2011anti}, the legitimate devices try to avoid the jamming signals by first detecting the jamming frequency range and then switching their operating frequency to another channel (within their operating frequency band). The direct sequence spread spectrum (DSSS) technique \cite{popper2010anti,liu2010randomized,simon2002spread}, on the other hand,  uses pseudo-random noise (PRN) codes to encode and spread the legitimate data signal to a much larger bandwidth, hence avoiding narrow-band jamming. Moreover, using the DSSS technique, the encoded data signals become noise-like and have low average power, making it challenging for the jammer to detect and interfere. However, when the operating frequency band of the legitimate devices is known to the jammers, they can perform wide-band jamming (i.e., given a sufficient jamming power) to void the functionality of the spread spectrum techniques.

Instead of ``dodging'' (i.e., by DSSS) or ``escaping'' (i.e., by FHSS), it has been shown that legitimate devices can also leverage jamming signals for data transmission. In \cite{van2019jam}, using the backscattering and energy harvesting, the legitimate transmitter can choose to either adapt its transmission rate, backscatter the jamming signals, harvest energy (from the jamming signals), or stay idle. In this way, the legitimate transmitter can mitigate the impact of the jamming signals and even utilize jamming energy for its communication. However, this approach requires additional hardware component, which is not always available, especially in mobile equipment. Moreover, backscattering communication is limited in its transmission range and throughput. 

Another popular anti-jamming solution is to estimate the jamming channels characteristics, such as their nullspace \cite{doukopoulos2008fast}, their projection \cite{do2017jamming}, their ratios \cite{yan2016jamming}, and then derive corresponding filters to suppress the jamming signals. These techniques merely require one degree-of-freedom to deal with each jammer, thus are more efficient than the aforementioned AoA-based approach. However, most of the existing works, e.g., \cite{do2017jamming,doukopoulos2008fast,yan2016jamming}, overlook the impact of the time-varying correlations between jamming signals on the nullification process. 
The authors of \cite{hoang2021suppression} have recently proved that by deliberately varying the correlations among jamming signals, attackers can effectively vary the jamming channels (referred to as the ``virtual change'') and thus their nullspace, even when the physical channels remain unchanged. This makes the estimated beam-forming matrix acquired from the estimated nullspace of the jamming channel outdated/incapable of nullifying the jamming signals. 

To deal with this ``virtual change'' in the jamming channels, one can continuously monitor the residual jamming signals (after applying the nullification technique) and then heuristically adjust the estimated beam-forming matrix when the residual surpasses a given threshold. However, such a jamming residual monitoring process incurs additional system overhead, thus significantly reducing the spectral efficiency. This gives rise to a more challenging problem on optimizing the duration of the nullspace estimation and the data transmission phases. A longer nullspace estimation phase may result in a lower jamming residual but then a shorter data transmission phase. This optimization problem is similar to the training phase duration optimization in MIMO communications (e.g.,\cite{hassibi2003much,wang2007performance}) as both errors (in channel estimation) lead to unwanted interference to the legitimate signals, thus reducing the post-equalization SINR and consequently the spectral efficiency. However, unlike \cite{hassibi2003much,wang2007performance} in which the training signals and the channel estimator can be designed to quantify the mean and variance of the channel estimation error, the jamming signals in our scenario are totally controlled by the jammers. This makes most training-window optimization approaches, e.g., \cite{hassibi2003much,wang2007performance} inapplicable. Note that suppressing jamming signals with an outdated estimated beam-forming matrix can result in even lower spectral efficiency than not using the method. This is because the jamming nullification requires legitimate users to sacrifice antennas/degree-of-freedoms that would be otherwise used for data transmission.

Given the above, this article proposes a solution to nullify multiple correlated jammers whose correlation is unknown and time-varying. To this end, we first systematically formulate the optimization problem of the nullspace estimation and data transmission phases. Even ignoring the unknown strategy of the jammers and the challenging nullspace estimation process, the resulting problem is an integer programming problem, hence intractable to obtain its optimal solution. In practice, as aforementioned, the jammers can deliberately vary the correlation range, making jamming nullification even more challenging. To deal with such uncertainty and incomplete information as well as to circumvent the intractability of the above conventional optimization problem, we reformulate the problem using a partially observable semi-Markov decision process (POSMDP). Then, we design a deep dueling Q-learning based technique to quickly obtain the optimal policy for the legitimate devices. The proposed technique does not require the legitimate devices to constantly monitor the residual jamming signals (and then update the estimated beam-forming matrix), and only costs a single degree-of-freedom to nullify each jammer, even with an unknown and time-varying correlated jamming strategy. At the beginning of the neural network in the proposed technique, we use the Long Short-Term Memory (LSTM) layer to capture the time-varying characteristic of the correlations and represent them by the output of the LSTM layer. The output of the LSTM layer is then used by a deep dueling neural network structure\cite{wang2016dueling}, which allows the proposed technique to obtain the optimal solution much faster than the conventional Q-learning, hence very effective in dealing with the time-varying jamming strategy of jammers. Simulation results show that the resulting spectral efficiency is much higher than that of other methods and close to that of the perfect jamming nullification case (for moderate jamming power). The proposed approach also yields a lower outage probability than the existing ones. The major contributions of the paper are as follows.

\begin{itemize}
	\item Derive the spectral efficiency bounds of each communication link using the estimated beam-forming matrix in the presence of jamming signals.
	\item Demonstrate that using an incorrectly estimated beam-forming matrix (i.e., resulting from the ``virtual change'' in the jamming channel caused by the time-varying correlations) can waste the receiver's degree-of-freedoms without achieving effective jamming suppression, corresponding to the lower bound spectral efficiency. Note that this lower-bound spectral efficiency is even lower than that of the case without using the estimated beam-forming matrix. 
	\item Formulate an optimization problem for the duration of the nullspace estimation and data transmission phases. The objective is to maximize the spectral efficiency and minimize the outage probability.
	\item Develop a partially observable semi-Markov decision process (POSMDP) framework to capture the dynamic and uncertainty of the jamming strategy and jamming channels. We then design a deep dueling Q-learning technique to obtain the optimal durations for the nullspace estimation and data transmission phases to maximize the system's spectral efficiency.
	\item Carry out intensive simulations to evaluate the performance of the proposed framework and compare it with start-of-the-art jamming suppression methods and the upper bound where the jamming is perfectly nullified.
\end{itemize}

The organization of the paper is as follows. Section \ref{Sec:System_model} describes the system model. Section \ref{Sec:Problem_formulation} formulates the problem being investigated. The deep dueling Q-learning technique to solve the stated problem is presented in Section \ref{Sec:Dueling}. The simulation results are given in Section \ref{Sec:Simulation Results}. Finally, the conclusions are drawn in Section VI.
\vspace{-0.5cm}
\section{System Model}
\label{Sec:System_model}
\vspace{-0.5cm}
\subsection{Network Model}
\begin{figure}[t]
	\centering
	\includegraphics[width=0.45\linewidth]{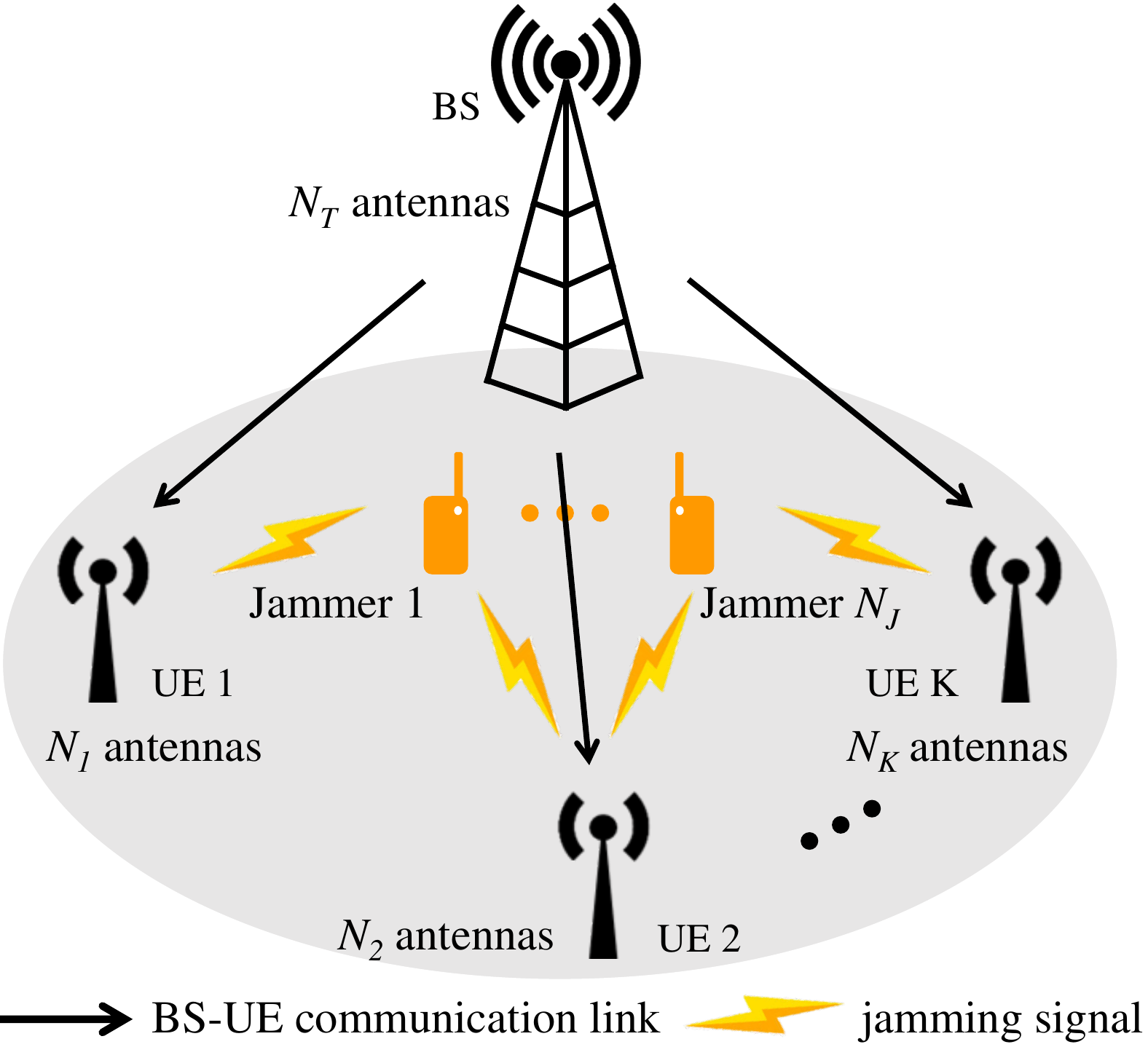}
	\vspace{0.2cm}
	\caption{BS-UE communication system interfered by proactive jammers.}
	\label{fig:block}
		\vspace{-0.5cm}
\end{figure}
We consider a multi-user multiple-input multiple-output (MU-MIMO) downlink system with one base station (BS) and $K$ user equipment (UEs) as demonstrated in Fig. \ref{fig:block}. The BS and each $k$th UE have ULA structures with $N_{\rm T}$ and $N_k$ antennas, respectively. The BS-UEs communication system is jammed/interfered by $N_{\rm J}$ single-antenna proactive jammers. Note that the techniques and results obtained in this paper also capture the case with multi-antenna jammers that can be considered as multiple single-antenna jammers. For either case, we assume the total number of jamming antennas is $N_{\rm J}$.
\vspace{-0.5cm}
\subsection{Signal Model}
The signal received at the $k$th UE can be expressed by
\begin{align}\label{eq:rec_sig_kth_user}
{\bf{y}}_{k}=\sqrt{P_{\rm T}}{\bf{H}}_k{\bf{P}}_k{\bf{x}}_k+\sqrt{P_{\rm T}}{\bf{H}}_k\sum^{K}_{l\neq k}{\bf{P}}_l{\bf{x}}_l+{\bf{Z}}_{k}{\bf{x}}_{\rm J}+{\bf{w}},
\\[-1 cm]\nonumber
\end{align}
where $P_{\rm T}$ is the transmitted power from the BS, ${\mathbf{H}}_k \in \mathbb{C}^{N_k \times N_{\rm T}}$ denotes the BS-$k$th UE channel, ${\mathbf{P}}_k\in \mathbb{C}^{N_T \times M_{k}}$ is the precoder applied at the BS for the $k$th UE, $M_{k}$ is the number of independent streams for the $k$th UE,  ${\bf{x}}_k\in\mathbb{C}^{M_{k}\times 1}$ denotes the signal transmitted from the BS to the $k$th UE, ${\bf{x}}_{\rm J}=[{\bf{x}}_{{\rm J}1};{\bf{x}}_{{\rm J}2};...{\bf{x}}_{{\rm J}N_{\rm J}}]\in\mathbb{C}^{N_{\rm J} \times 1}$ is the transmitted jamming signal,  ${\bf{Z}}_{k}\in \mathbb{C}^{N_k \times N_{\rm J}}$ is the jammers-$k$th UE channel, and ${{\bf{w}}}\in\mathbb{C}^{N_k \times 1}$ is complex noise. The elements of ${{\bf{w}}}$ are assumed to be zero-mean circularly-symmetric complex Gaussian random variables (i.e., ${\bf{w}}\sim\mathcal{CN}({\mathbf{0}},\sigma_w^2\textbf{I}_{N_k})$, where $\textbf{I}_{N_k}$ denotes the identity matrix of size ${N_k}$, and $\sigma_w^2$ is the noise variance). Likewise, we assume ${\bf{x}}_k\sim\mathcal{CN}({\mathbf{0}},\textbf{I}_{M_{k}})$. The precoding matrix ${\mathbf{P}}_k$ is normalized to meet the power constraint, such that $||{\mathbf{P}}_k||_F=1$, where $||.||_F$ denotes the Frobenius norm of a matrix.
\vspace{-0.5cm}
\subsection{Jamming Signal Model}
\label{jamming_model}
Similar to \cite{hoang2021suppression}, we assume ${\bf{x}}_{\rm J}\sim{\mathcal{CP}}(\bm{{\mu}}_{\rm J},{\bm{\Sigma}}_{\rm J})$, where $\mathcal{P}$ denotes a distribution function, $\bm{{\mu}}_{\rm J}$ and ${\bm{\Sigma}}_{\rm J}$ are the mean and covariance matrix of ${\bf{x}}_{\rm J}$. Note that the jamming strategy, $\mathcal{P}$, $\bm{{\mu}}_{\rm J}$ and ${\bm{\Sigma}}_{\rm J}$ are unknown to the BS and the UEs.

Let $\rho_{ij}$ be the complex Pearson correlation between the transmitted jamming signals from the $i$th and the $j$th jammer. The value of $\rho_{ij}$ can be expressed by \cite[Ch.~4]{schreier2010statistical}
\begin{align}
\label{Pearson's formula}
\rho_{ij}=\frac{{\mathbb{E}}({\bf{X}}_{{\rm J}i}{\bf{X}}^H_{{\rm J}j})}{\sigma_{{\rm J}_{i}}\sigma_{{\rm J}_{j}}},
\\[-1 cm]\nonumber
\end{align}
where ${\bf{X}}_{{\rm J}j}$ denotes a sample set of ${\bf{x}}_{{\rm J}j}$, and $\sigma_{{\rm J}_{j}}^2$ is the variance of the $j$th transmitted jamming signal. The covariance matrix $\mathbf{\Sigma}_{\rm J}$ can be expressed by
\begin{align}
\label{covariance_transmitted_signal}
\mathbf{\Sigma}_{\rm J}=\begin{bmatrix}\sigma_{{\rm J}_{1}}^2&\rho_{12}\sigma_{{\rm J}_{1}\sigma_{2}}&...&\rho_{1N_{\rm J}}\sigma_{{\rm J}_{1}}\sigma_{{\rm J}_{N_{\rm J}}}\\\rho_{12}^*\sigma_{{\rm J}_{1}}\sigma_{{\rm J}_{2}}&\sigma_{{\rm J}_{2}}^2&...&\rho_{2N_{\rm J}}\sigma_{{\rm J}_{2}}\sigma_{{\rm J}_{N_{\rm J}}}\\...&...&...&...\\\rho_{1N_{\rm J}}^*\sigma_{{\rm J}_{1}}\sigma_{{\rm J}_{N_{\rm J}}}&\rho_{2N_{\rm J}}^*\sigma_{{\rm J}_{2}}\sigma_{{\rm J}_{N_{\rm J}}}&...&\sigma_{{\rm J}_{N_{\rm J}}}^2\end{bmatrix}.
\\[-1 cm]\nonumber
\end{align}

As mentioned in the introduction and will be described in more detail in Subsection \ref{subsec:impact_correlation}, the time-varying correlations between transmitted jamming signals create a ``virtual change'' in the jamming channel, even when the physical channels stay unchanged. Based on the behavior of the ``virtual change''in the jamming channel, the jammers can disable the functionality of the conventional jamming suppression techniques by deliberately varying the correlations amongst jamming signals. Formally, the correlations between the transmitted jamming signals are controlled by the jammers using the formula
\begin{align}
\label{eqn:corr}
\rho_{ij}(t)=\mathcal{J}(i,j,t), \forall i\neq j\in(1,2,...,N_{\rm J}),
\\[-1 cm]\nonumber
\end{align}
where $\mathcal{J}$ is a function unknown to the UEs and the BS.
\vspace{-0.5cm}
\subsection{Channel Model}
The BS-$k$th UE channel can be given by
\begin{align}
\label{eqn:BS-UE_channel}
{\bf H}_{k}&=\frac{1}{\sqrt{\eta_{k}}}\sum_{p=1}^{N^{\rm {p}}_{k}} \alpha_{k,p} {\bf a}({\phi}_{k,p}^{\rm a}){\bf a}({\phi}_{k,p}^{\rm d})^T,
\\[-1 cm]\nonumber
\end{align}
where ${\eta_{k}}$ denotes the large-scale path-loss of the BS-$k$th UE channel, $\alpha_{k,p}$ is the complex path gain, $N^{\rm {p}}_{k}$ is the total number of propagation paths, ${\phi}_{k,p}^{\rm a}$ and ${\phi}_{k,p}^{\rm d}$ are the AoA and angle of departure (AoD) of the $p$th path for the $k$th UE, respectively, and ${\bf a}({\phi}_{k,p}^{\rm a})$ and ${\bf a}({\phi}_{k,p}^{\rm d})$ are the steering vectors corresponding to ${\phi}_{k,p}^{\rm a}$ and ${\phi}_{k,p}^{\rm d}$, respectively. The steering vectors can be expressed by\cite[Ch.~3]{chen2010introduction}
\begin{align}
{\bf a}({\phi}_{k,p}^{\rm a})&=\left[1,e^{-j\frac{2\pi d_k}{\lambda}\sin({\phi}_{k,p}^{\rm a})},...,e^{-j\frac{2\pi d_k}{\lambda}(N_k-1)\sin({\phi}_{k,p}^{\rm a})}\right]^T,\nonumber\\
{\bf a}({\phi}_{k,p}^{\rm d})&=\left[1,e^{-j\frac{2\pi d_{\rm T}}{\lambda}\sin({\phi}_{k,p}^{\rm d})},...,e^{-j\frac{2\pi d_{\rm T}}{\lambda}(N_{\rm T}-1)\sin({\phi}_{k,p}^{\rm d})}\right]^T,\nonumber
\\[-1 cm]\nonumber
\end{align}
where $d_k$ and $d_{\rm T}$ are the antenna element spacing at the $k$th UE and the BS, respectively, and $\lambda$ is the carrier's wavelength. The AoA ${\phi}_{k,p}^{\rm a}$ is assumed to be uniformly distributed over $[0,2\pi]$\cite{prabhu2002simulation}. We further assume that $N^{\rm {p}}_{k}$ is sufficiently large (e.g., $N^{\rm {p}}_{k}\geq$8), such that the elements of ${\bf H}_{k}$ are zero-mean circularly-symmetric complex Gaussian random variables (i.e., ${\bf H}_{k}\sim\mathcal{CN}({\mathbf{0}},1/\sqrt{\eta_{k}})$).

Similarly, the $j$th jammer-$k$th UE channel is given as
\begin{align}
\label{eqn:inter-UE_channel}
{\bf Z}_{k,j}&=\frac{1}{\sqrt{\eta_{k,j}}}\sum_{p=1}^{N^{\rm p}_{k,j}} \alpha_{k,j,p} {\bf a}({\phi}_{k,j,p}^{\rm a}),
\\[-1 cm]\nonumber
\end{align}
with ${\eta_{k,j}}$, $N^{\rm p}_{k,j}$, $\alpha_{k,j,p}$, ${\phi}_{k,j,p}^{\rm a}$, and ${\bf a}({\phi}_{k,j,p}^{\rm a})$ defined in the same way to ${\eta_{k}}$, $N^{\rm {p}}_{k}$, $\alpha_{k,p}$, ${\phi}_{k,p}^{\rm a}$, and ${\bf a}({\phi}_{k,p}^{\rm a})$ in Eq. (\ref{eqn:BS-UE_channel}), respectively. The COST 231 Hata model \cite[Ch.~4]{damosso1999digital} is used to model the large-scale path-losses ${\eta_{k}}$ and ${\eta_{k,j}}$. For the multipath fadings (i.e., expressed by the summations in Eq. (\ref{eqn:BS-UE_channel}) and Eq. (\ref{eqn:inter-UE_channel})), without loss of generality, we adopt the flat fast fading Rayleigh model in \cite{xiao2006novel}. The values of related parameters are specified in Section \ref{Sec:Simulation Results}.
{\setlength{\extrarowheight}{3pt}%
	\begin{table}[t]
		\small	
		\caption{\textsc{Notation and Symbols.}}
		\begin{tabular}{|P{1.2cm}|p{6cm}|P{1.2cm}|p{6cm}|}
			\hline 
			\textbf{Notation} & \textbf{Description}&\textbf{Notation}&\textbf{Description}\\ 
			\hline
			${\hat{\bf{F}}}_{k}$&Estimated beam-forming matrix for the $k$th UE.&		$|.|$&Modulus of complex number or dimension of space.\\
			\hline
			$N^{\rm e}$,\break $N^{\rm d}$& Number of samples in the nullspace estimation and data transmission phases, respectively.&${\bf{\mathcal{N}}^{\rm e}}$, \break ${\bf{\mathcal{N}}^{\rm d}}$& Sets of candidates for $N^{\rm e}$ and $N^{\rm d}$, respectively.\\
			\hline
			$\delta_{k,m}$,\break $\hat{\delta}_{k,m}$,\break $\bar{\delta}_{k}$, \break $\delta_{\rm {min}}$& Post-equalization signal-to-interference-plus-noise ratio (SINR) of the $m$th stream for the $k$th UE, estimate  of $\delta_{k,m}$, average of $\hat{\delta}_{k,m}$, and minimum required post-equalization SINR, respectively.&			$\mathcal{S}$, \break$\hat{\mathcal{S}}$, \break $\mathcal{A}$, \break $r$&\break State space, approximate state space, action space, and immediate reward, respectively.\\
			\hline
			$\Lambda_{k,l}$\break$\bar{\Lambda}_{l}$ & The $l$th largest singular value of ${\bf{R}}_{{\rm J}_{k}}^{\rm e}$ and the average of $\Lambda_{k,l}$, respectively.&$\mathcal{Q}$, \break $\hat{\mathcal{Q}}$&Q-network and target Q-network, respectively.\\
			\hline
			$s[n],\break \hat{s}[n] \break a[n],\break r[n]$&\break State, approximate state, action, and immediate reward at the $n$th epoch, respectively.&$(.)^{-1}$,\break  $(.)^*$,\break $(.)^T$,\break $(.)^H$& \break The inverse, transpose, conjugate, and Hermitian transpose matrix operations, respectively.\\
			\hline 
			$\mathcal{V}(s)$,\break$Q(s,a)$\break $\mathcal{G}({s},a)$&Value function, state-action value function, and action advantages function, respectively.&${\mbox{Var}}(.)$\break ${\mbox{Cov}}(.)$&The variance and covariance, respectively.\\
			\hline
		\end{tabular}
		\vspace{-1cm}
		\label{Tab:notation}
\end{table}}

For reference purposes, important notations and symbols are given in Table. \ref{Tab:notation}. We employ the superscripts ``e'' and ``d'' to denote symbols in the nullspace estimation and data transmission phases, respectively. For example, $\mathbf{\Sigma}_{\rm J}^{\rm e}$ and $\mathbf{\Sigma}_{\rm J}^{\rm d}$ represent the values of $\mathbf{\Sigma}_{\rm J}$ in the nullspace estimation phase and data transmission phase, respectively. $\hat{(.)}$ denotes the estimated or approximate value. For example, the estimated value of ${\bf G}_k$ (introduced in Section \ref{Sec:Problem_formulation}) is denoted by $\Gk$. 
\vspace{-0.5cm}
\section{Problem Formulation}
\label{Sec:Problem_formulation}
We first describe the communication protocol and the estimated beam-forming matrix employed to suppress the jamming signals. We then briefly analyze the impacts of the time-varying correlations among jamming signals on the jamming suppression process (using the estimated beam-forming matrix). Next, we derive the upper and lower bounds for the spectral efficiency of each BS-UE communication link that employs the estimated beam-forming matrix to suppress the jamming signals. Finally, we mathematically state the problem. To highlight the impact of the time-varying correlations on designing the estimated beam-forming matrix, we assume that the BS-UEs and the jamming channels follow a block-fading model with coherence time $T^{\rm c}$, corresponding to $N^{\rm c}$ samples. We further assume that the nullspace estimation, preamble, and data transmission phases of the communication protocol (described below) are performed within the interval $T^{\rm c}$, such that $N^{\rm e}+N^{\rm p}+N^{\rm d}<N^{\rm c}$, where $N^{\rm e},N^{\rm p},N^{\rm d}$ are the number of samples of the nullspace estimation, preamble, and data transmission phase, respectively.
\vspace{-0.5cm}
\subsection{Communication Protocol}
\label{subsec:communication protocol}
\begin{figure}[t]
	\centering
	\includegraphics[width=0.45\linewidth]{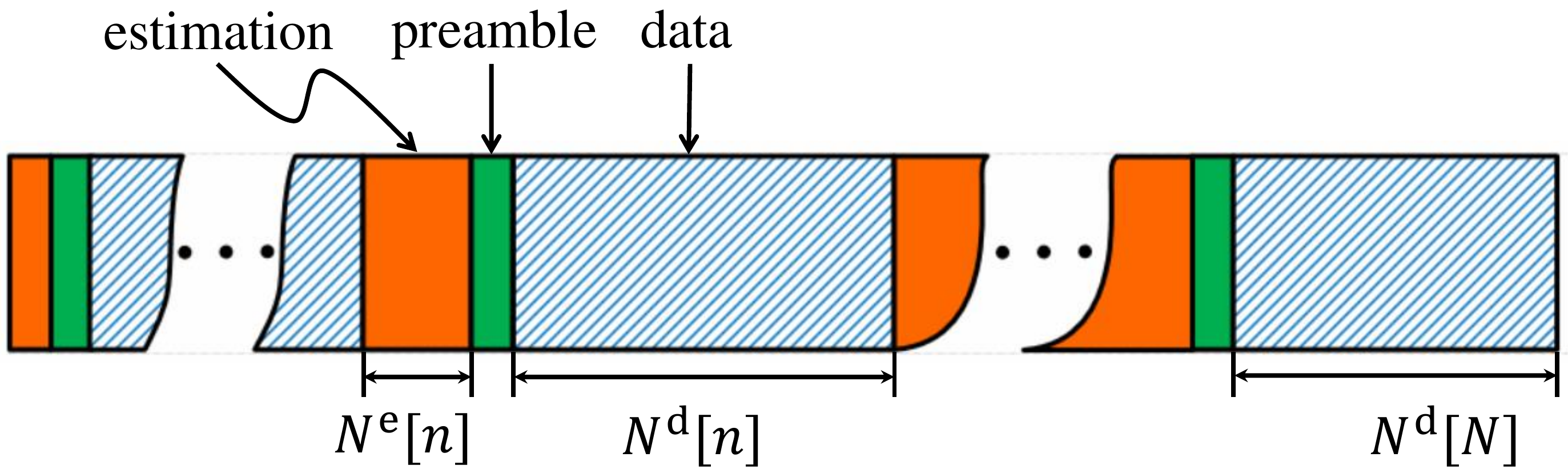}
	\vspace{0.2cm}
	\caption{Communication protocol for jamming suppression.}
	\label{fig:frame}
	\vspace{-0.5cm}
\end{figure}

Fig. \ref{fig:frame} illustrates the communication protocol for the jamming/interference nullification purpose \cite{leost2012interference}. As shown, each frame is comprised of three phases: nullspace estimation, preamble, and data transmission.
\begin{itemize}
	\item During the nullspace estimation phase, which lasts for $N^{\rm e}$ samples, the beam-forming matrix ${\hat{\bf{F}}}_{k}$ that is used to suppress the jamming signals, is estimated. Let ${\bf{F}}_{k}$ denote the actual value of ${\hat{\bf{F}}}_{k}$. ${\bf G}_k\in \mathbb{C}^{(N_k-N_{\rm J}) \times N_k}$ is a matrix whose rows form an orthonormal basis for the left nullspace\cite[Ch.~2]{strang2016introduction} of the received jamming signals, and let $\Gk$ denotes the estimated value of ${\bf G}_k$. The estimated beam-forming matrix ${\hat{\bf{F}}}_{k}$ is designed by choosing its rows from the rows of $\Gk$. Therefore, let $B_k$ be the row number of ${\hat{\bf{F}}}_{k}$, one has $B_k\leq(N_k-N_{\rm J})$. To leverage all $(N_k-N_{\rm J})$ remaining degree-of-freedoms (after jamming suppression) for BS-UE signal multiplexing, we set ${\hat{\bf{F}}}_{k}=\Gk$ by letting $B_k=(N_k-N_{\rm J})$. Similar to \cite{doukopoulos2008fast}, $\Gk$ can be estimated using the singular value decomposition (SVD) as follows.
		
	Let $\bf{x}_{\rm J}^{\rm e}$ and ${\bf{y}}_{{\rm J}_{k}}^{\rm e}$ denote the values of $\bf{x}_{\rm J}$ and ${\bf{y}}_{k}$ during the nullspace estimation phase, respectively. Let ${\bf{Y}}_{{\rm J}_{k}}^{\rm e}$ and ${\bf{R}}_{{\rm J}_{k}}^{\rm e}$ be a sample set and the corresponding covariance with $N^{\rm e}$ samples of ${\bf{y}}_{{\rm J}_{k}}^{\rm e}$. During the nullspace estimation phase, the BS does not send any signal, hence we have,
		\begin{align}
		{\bf{y}}_{{\rm J}_{k}}^{\rm e}={\bf{Z}}_{k}{\bf{x}}_{\rm J}^{\rm e}+{\bf{w}},\qquad{\bf{R}}_{{\rm J}_{k}}^{\rm e}=\frac{1}{{N}^{\rm{e}}}{\bf{Y}}_{{\rm J}_{k}}^{\rm e}({\bf{Y}}_{{\rm J}_{k}}^{\rm e})^H.
		\\[-1 cm]\nonumber
		\end{align}
We have $\Gk=({\bf U}_{\rm w})^{{H}},$ where ${\bf U}_{\rm w}$ is extracted from the SVD of ${\bf{R}}_{{\rm J}_{k}}^{\rm e}$,
		\begin{align}\label{SVD_covariance}
		{\bf{R}}_{{\rm J}_{k}}^{\rm e}=[{\bf U}_{\rm s}\;{\bf U}_{\rm w}]\begin{bmatrix}{\bf \Lambda}_{\rm s} &\bf{0}\\\bf{0}&{\bf \Lambda}_{\rm w}\end{bmatrix}\begin{bmatrix}({\bf U}_{\rm s})^{H}\\({\bf U}_{\rm w})^{H}\end{bmatrix}.
		\\[-1 cm]\nonumber
		\end{align}
	\item During the preamble phase, which lasts for $N^{\rm p}$ samples, the jamming signals are nullified by multiplying Eq. (\ref{eq:rec_sig_kth_user}) with ${\hat{\bf{F}}}_{k}$, and the BS-UE equivalent channel $\tilde{{\bf{H}}}_k$ (i.e., $\tilde{{\bf{H}}}_k={\hat{\bf{F}}}_{k}{\bf{H}}_k{\bf{P}}_k$) is estimated. The estimation of $\tilde{{\bf{H}}}_k$ can be performed using pilot signals and a channel estimator, such as the minimum mean-square error (MMSE) or least-square (LS) technique.
	
	\item During the data transmission phase, which lasts for $N^{\rm d}$ samples, the BS send data to UEs. Let $\delta_{k,m}[n]$ denote the post-equalization signal-to-interference-plus-noise ratio (SINR) of the $m$th stream for the $k$th UE during the data transmission phase of the $n$th frame. As will be demonstrated in Appendix \ref{appen_theorem2}, $\delta_{k,m}[n]$ can be given by 
	\begin{align}
	\label{snr}
	\delta_{k,m}[n]=\frac{P_{\rm T}}{{\mbox{Var}}\{[{{\bf{A}}_k^{\rm {ZF}}}(\delta{\bf{F}}_{k}{\bf{Z}}_{k}{\bf{x}}_{\rm J}+{\hat{\bf{F}}}_{k}{\bf{w}})]_m[n]\}}.\quad\qquad\qquad\qquad\qquad\qquad
	\\[-1 cm]\nonumber
	\end{align}
	where $\mbox{Var}(.)$ denotes the variance, ${{\bf{A}}_k^{\rm {ZF}}}$ is the zero-forcing equalizer, $(.)_{m}$ denotes the $m$th elements of a vector, $\delta{\bf{F}}_{k}=\hat{\bf{F}}_{k}-{\bf{F}}_{k}$ is the estimation error of ${\bf{F}}_{k}$.
	
	Let $C_{k,m}[n]$, and $C_{k,m}^{\rm {eff}}[n]$ denote the corresponding spectral efficiency and effective spectral efficiency, respectively. We have
	\begin{align}
	C_{k,m}[n]&=\log_{2}\{1+\delta_{k,m}[n]\},\qquad C_{k,m}^{\rm {eff}}[n]=\mu[n]C_{k,m}[n],\qquad\qquad\qquad\;\label{spectral_effective}
	\end{align}
	where
	\begin{align}
	\mu[n]=\frac{N^{\rm d}[n]}{N^{\rm e}[n]+N^{\rm p}[n]+N^{\rm d}[n]}\qquad\qquad\qquad\qquad\qquad\qquad\qquad\label{eff_ratio}
	\\[-1 cm]\nonumber
	\end{align}
 is the data transmission phase fraction over the whole frame.
\end{itemize}
\vspace{-0.5cm}
\subsection{Impact of the Time-varying Correlations on Jamming Suppression}
\label{subsec:impact_correlation}
The estimated beam-forming matrix ${\hat{\bf{F}}}_{k}$ described in the previous subsection is derived from $\Gk$, whose rows form an orthonormal basis for the left nullspace of the jamming signal. Within the jamming channel's coherence time \cite{manolakos2012blind}, ${\hat{\bf{F}}}_{k}$ is capable of nullifying the jamming signals. Nevertheless, as shown in \cite{hoang2021suppression}, when the correlations between transmitted jamming signals are time-varying, they create a ``virtual change'' in the jamming channel, hence degrading the estimation accuracy of $\Gk$ and making ${\hat{\bf{F}}}_{k}$ unable to suppress the jamming signals. Specifically, let  $\rho_{ij}^{\rm e}$ and $\rho_{ij}^{\rm d}$ be $\rho_{ij}$ values in the nullspace estimation phase and data transmission phase, respectively. Similarly, let $\mathbf{\Sigma}_{\rm J}^{\rm e}$ and $\mathbf{\Sigma}_{\rm J}^{\rm d}$ denote $\mathbf{\Sigma}_{\rm J}$ values in these two phases. The time-varying values of $\rho_{ij}$ from $\rho_{ij}^{\rm e}$ to $\rho_{ij}^{\rm d}$ effectively change the jamming channels from ${\bf Z}_k$ to (${\bf Z}_k{\bf{D}}$) where: 
\begin{align}
\label{vir_change}
{\bf{D}}&={\bf{V}^{\rm d}}\sqrt{{\bf{S}}^{\rm d}({\bf{S}}^{\rm e})^{-1}}({\bf{V}^{\rm e}})^{H},\qquad\qquad\quad\;
\\[-1 cm]\nonumber
\end{align} in which
\begin{align}
\label{jamsig_svd}
\mathbf{\Sigma}_{\rm J}^{\rm d}={\bf{V}^{\rm d}}{\bf{S}}^{\rm d}({\bf{V}^{\rm d}})^{H}\text{ and } \mathbf{\Sigma}_{\rm J}^{\rm e}={\bf{V}^{\rm e}}{\bf{S}}^{\rm e}({\bf{V}^{\rm e}})^{H}
\\[-1 cm]\nonumber
\end{align}
are the SVD of $\mathbf{\Sigma}_{\rm J}^{\rm d}$ and $\mathbf{\Sigma}_{\rm J}^{\rm e}$, respectively.

Interesting intuitions about the impact of time-varying correlations on the jamming suppression capability of ${\hat{\bf{F}}}_{k}$ can be obtained by examining the behavior of the ``virtual change'' factor ${\bf{D}}$ in the jamming channel. First, when the correlations are unchanged over time, from Eq. (\ref{covariance_transmitted_signal}) and Eq. (\ref{jamsig_svd}), $\mathbf{\Sigma}_{\rm J}^{\rm d}=\mathbf{\Sigma}_{\rm J}^{\rm e}$, ${\bf{S}}^{\rm d}={\bf{S}}^{\rm e}, {\bf{V}^{\rm d}}={\bf{V}^{\rm e}}$, and hence, from Eq. (\ref{vir_change}), ${\bf{D}}={\bf{I}}$. Therefore, when the correlations are unchanged over time, there is no ``virtual change'' in the jamming channel. Therefore, in this case, within the jamming channel's coherence time, ${\hat{\bf{F}}}_{k}$ derived from $\Gk$ can be utilized to suppress the jamming signals regardless of the (fixed) correlation values. Second, when the correlations are time-varying, the behavior of elements of the non-identity ``virtual change'' matrix ${\bf{D}}$ is described by Corollary $1.1$  \cite{hoang2021suppression} below.

\textit{\bf{Corollary 1.1}:} When $|\rho_{ij}^{\rm e}|\to1$ and $\rho_{ij}^{\rm d}\neq\rho_{ij}^{\rm e}$, the elements of the ``virtual change'' factor ${\bf{D}}$ approach infinity.

The elements of ${\bf{D}}$ approach infinity when $|\rho_{ij}^{\rm e}|\to1$. As a result, the estimation $\Gk$ is not accurate, and thus ${\hat{\bf{F}}}_{k}$ is not close to ${\bf{F}}_{k}$. In this case, using ${\hat{\bf{F}}}_{k}$ leads to significant jamming residual and hence a low post-equalization SINR at the receiver in the data transmission phase. The jammers then can use the jamming signals model in Eq. (\ref{eqn:corr}) to maximize the jamming impact by degrading the estimation accuracy of ${\hat{\bf{F}}}_{k}$. Theorem $2$ below analyzes the effectiveness of ${\hat{\bf{F}}}_{k}$ in nullifying the jamming signals by examining the lower and upper spectral efficiency bounds of each BS-UE communication link. For illustration purposes, we assume that the zero-forcing equalization is used at the UE receiver.

\textit{\bf{Theorem 2}:} The spectral efficiency of the $m$th stream for the $k$th UE is bounded by
\begin{align}
	C_{k,m}^{\rm lb}\leq C_{k,m}\leq C_{k,m}^{\rm ub},
\\[-1 cm]\nonumber	
\end{align}
where $C_{k,m}^{\rm lb}$ and $C_{k,m}^{\rm ub}$ are the lower and upper bounds of $C_{k,m}$
\begin{align}
C_{k,m}^{\rm lb}=&\log_{2}\big[1+\frac{{P_{\rm T}(N_k-N_J-M_{k})}}{\eta_{k}(\sigma_w^2+\sum_{j=1}^{N_J}{\eta_{k,j}}\sigma_{{\rm J}_{j}}^2)}\big],\label{capacit_lb}\\
C_{k,m}^{\rm ub}=&\log_{2}\big[1+\frac{{P_{\rm T}(N_k-N_J-M_{k})}}{\eta_{k}\sigma_w^2}\big].\label{capacit_ub}
\\[-1 cm]\nonumber
\end{align}
\textit{Proof:} The proof is given in Appendix \ref{appen_theorem2}. \hfill $\blacksquare$

Without nullifying/suppressing the jamming signals (or without beam-forming), the corresponding spectral efficiency, referred to as $C_{k,m}^{\rm wbf}$, is
\begin{align}
\label{capacity_wbf}
C_{k,m}^{\rm wbf}=&\log_{2}\big[1+\frac{{P_{\rm T}(N_k-M_{k})}}{\eta_{k}(\sigma_w^2+\sum_{j=1}^{N_J}{\eta_{k,j}}\sigma_{{\rm J}_{j}}^2)}\big].
\\[-1 cm]\nonumber
\end{align}
As can be seen, the use of the estimated beam-forming matrix ${\hat{\bf{F}}}_{k}$ does not always guarantee better spectral efficiency than not using it, i.e., $C_{k,m}^{\rm wbf}> C_{k,m}^{\rm lb}$. It is because when using ${\hat{\bf{F}}}_{k}$, the UE receiver has to ``sacrifice'' $N_J$ degree-of-freedoms, as demonstrated by comparing the numerators of Eq. (\ref{capacit_lb}) and Eq. (\ref{capacit_ub}) to that of Eq. (\ref{capacity_wbf}). Especially, when the estimated beam-forming matrix ${\hat{\bf{F}}}_{k}$ is not accurate (e.g., because of the impact of the time-varying correlations as illustrated in Corollary $1.1$), its use can lead to the worst case with the lower-bound spectral efficiency. In this case, the UE receiver lost $N_J$ degree-of-freedoms without achieving any jamming nullification effect, leaving the denominator of Eq. (\ref{capacit_lb}) unchanged compared to that of Eq. (\ref{capacity_wbf}). On the other hand, when ${\hat{\bf{F}}}_{k}$ is estimated perfectly (i.e., ${\hat{\bf{F}}}_{k}={{\bf{F}}}_{k}$), the UE receiver can completely nullify the jamming signals, leaving only the noise in the denominator of Eq. (\ref{capacit_ub}), thereby achieving the upper-bound spectral efficiency. In the sequel, we aim to accurately estimate the beam-forming matrix ${\hat{\bf{F}}}_{k}$, even when the correlations are time-varying; thereby targeting the perfect beam-forming, with the spectral efficiency of each stream close to the upper bound given in Eq. (\ref{capacit_ub}).
\vspace{-0.5cm}
\subsection{Problem Formulation}
Given the above, one can maximize the spectral efficiency by continuously adapting the length of the nullspace estimation and data transmission phases (i.e., $N^{\rm e}$ and $N^{\rm d}$, respectively). Specifically, $N^{\rm e}$ and $N^{\rm d}$ can be tuned based on the following principles.

\begin{itemize}
	\item First, $N^{\rm e}$ and $N^{\rm d}$ are jointly optimized to guarantee that the estimation ${\hat{\bf{F}}}_{k}$ can be obtained when none of $|\rho_{ij}^{\rm e}|$ is close to $1$. As presented in Corollary $1.1$, when $|\rho_{ij}^{\rm e}|\to1$, the elements of ${\bf{D}}$ approach infinity, making the estimated beam-forming matrix ${\hat{\bf{F}}}_{k}$ unable to suppress the jamming signals, and hence resulting in a lower spectral efficiency.
	\item Second, by saving the time spent on monitoring the residual jamming signals as in \cite{hoang2021suppression} to update the estimated beam-forming matrix, the effective spectral efficiency of the system can be significantly improved, as demonstrated in Eq. (\ref{spectral_effective}). In fact, only when necessary, the system may increase the nullspace estimation time (by increasing $N^{\rm e}$) to average the correlations between transmitted jamming signals and avoid $|\rho_{ij}^{\rm e}|$ being close to $1$. 
	\item Third, by adjusting $N^{\rm d}$, the communication system can adapt to the change in the BS-UE channel condition. For example, when the channel coherence time decreases, the value of $N^{\rm d}$ should be decreased to maintain an acceptable post-equalization SINR level (e.g., above the required minimum post-equalization SINR, below which the UE is considered to be an outage). On the other hand, when the coherence time increases, the system can increase $N^{\rm d}$ to improve the communication phase fraction over the whole frame, hence increasing the effective spectral efficiency. 
\end{itemize}
The optimization of the durations of the nullspace estimation and data transmission phases can be formally stated as follows:
	\begin{align}
	\label{eqn:formulation_1}
	\max_{N^{\rm e},N^{\rm d}}\:\:&\lim_{N\to\infty}\Big\{\frac{1}{N}\sum_{n=1}^{N}\sum_{k=1}^{K}\sum_{m=1}^{M_{k}}\mu[n]\log_{2}\{1+\delta_{k,m}[n]\}\Big\}\\
	\textbf{s.t.}\qquad&\mu[n]=\frac{N^{\rm d}[n]}{N^{\rm e}[n]+N^{\rm p}[n]+N^{\rm d}[n]}\text{ as in (\ref{eff_ratio})},\nonumber\\
	&\delta_{k,m}[n]=\frac{P_{\rm T}}{{\mbox{Var}}\{[{{\bf{A}}_k^{\rm {ZF}}}(\delta{\bf{F}}_{k}{\bf{Z}}_{k}{\bf{x}}_{\rm J}+{\hat{\bf{F}}}_{k}{\bf{w}})]_m[n]\}}\nonumber\text{ as in (\ref{snr})},\quad\nonumber\\
	&\delta{\bf{F}}_{k}=\hat{\bf{F}}_{k}-{\bf{F}}_{k},\quad{{\bf{A}}_k^{\rm {ZF}}}=(\tilde{{\bf{H}}}_k^H\tilde{{\bf{H}}}_k)^{-1}\tilde{{\bf{H}}}_k^H,\nonumber\\
	&\tilde{{\bf{H}}}_k={\hat{\bf{F}}}_{k}{\bf{H}}_k{\bf{P}}_k,\quad\delta_{k,m}[n]\geq\delta_{\rm {min}},\nonumber
	\\[-1 cm]\nonumber
	\end{align}
where $N$ is the number of frames and $\delta_{\rm {min}}$ is the required minimum post-equalization SINR, below which the UE is considered to be an outage.

There are analogies between the estimation error of ${\bf{F}}_{k}$ and that of the BS-UE channel in MIMO communications (e.g.,\cite{hassibi2003much,wang2007performance}), because both errors lead to unwanted interference to the legitimate signals, thus reducing the post-equalization SINR and consequently the spectral efficiency. However, unlike \cite{hassibi2003much,wang2007performance} in which the training signals and the channel estimator can be designed to quantify the mean and variance of the channel estimation error, the jamming signals in our scenario are controlled by the jammers (i.e., described by Eq. (\ref{eqn:corr})). Therefore, the BS and UE do not have knowledge of the mean and variance of $\delta{\bf{F}}_{k}$. More importantly, to make jamming suppression even more challenging, the jammers can deliberately adjust the correlations controlling function $\mathcal{J}$, making the previous measurements no longer representative of the current jamming strategy. To deal with such incomplete information and uncertainty, in the next section, we reformulate the deep dueling Q-learning technique to solve the problem stated in Eq. (\ref{eqn:formulation_1}).
\vspace{-0.5cm}
\section{Deep Dueling Q-Learning Technique for Jamming Suppression}
\label{Sec:Dueling}
This section reformulate the problem (\ref{eqn:formulation_1}) using a partially observable semi-Markov decision process (POSMDP). We then design a deep dueling Q-learning based framework to tune the durations of the nullspace estimation and data transmission phases by obtaining the optimal policy for the underlying POSMDP process.
\vspace{-0.5cm}
\subsection{Partially Observable Semi-Markov Decision Process (POSMDP)}
A conventional MDP is defined by a tuple $(\mathcal{S},\mathcal{A},r)$, where $\mathcal{S}$, $\mathcal{A}$, and $r$ denote the state space, action space, and the reward function, respectively. An SMDP, on the other hand, retains the three components mentioned above and adds an additional component that is the $n$th decision epoch length, denoted by $t[n]$. In an MDP, the state transition occurs at regular time steps (and hence the decision epoch length $t[n]$ is excluded). On the other hand, the SMDP allows the state transition to occur at irregular time steps (i.e., different $t[n]$ for different epochs), facilitating the selections of $N^{\rm e}$ and $N^{\rm d}$ at irregular state transition times. 
\subsubsection{State Space}There are several essential factors to consider for maximizing the effective spectral efficiency of the system while avoiding outage. The first factor is the post-equalization SINR at the UEs during the previous data transmission phase. This is because the post-equalization SINR implicitly captures the BS-UE channel condition that affects the selection of $N^{\rm d}$. Specifically, when the post-equalization SINR in the previous data transmission phase is poor (e.g., below or close to $\delta_{\rm {min}}$), $N^{\rm d}$ value can be decreased to improve the post-equalization SINR. Hence, the system can avoid outages and increase spectral efficiency during the data transmission phase. However, $N^{\rm d}$ should not be incautiously decreased, as that reduces the fraction of the data transmission phase over the whole frame and thus reduces the effective spectral efficiency of the system. The second factor, as demonstrated in the previous section by Corollary 1.1, is the correlation between transmitted jamming signals in the nullspace estimation phase. This is because the correlations $\rho_{ij}^{\rm e}$ affect the magnitude of the ``virtual change'' factor ${\bf{D}}$ in the jamming channel, which directly affects the jamming nullification capability of ${\hat{\bf{F}}}_{k}$. Therefore, the system's state space can be defined as
\begin{align}
\label{eqn:State}
\mathcal{S}\delequal\big\{(\delta_{k,m},\rho_{ij}^{\rm e}):&\forall k\in (1,2,...,K),\nonumber\\
\forall m\in(1,2,...,M_{k}), &\forall i\neq j; i,j\in(1,2,...,N_{\rm J})
\big\}.
	\\[-1 cm]\nonumber
\end{align}
\subsubsection{Observation}
\begin{figure}[t]
	\centering
	\includegraphics[width=0.45\linewidth]{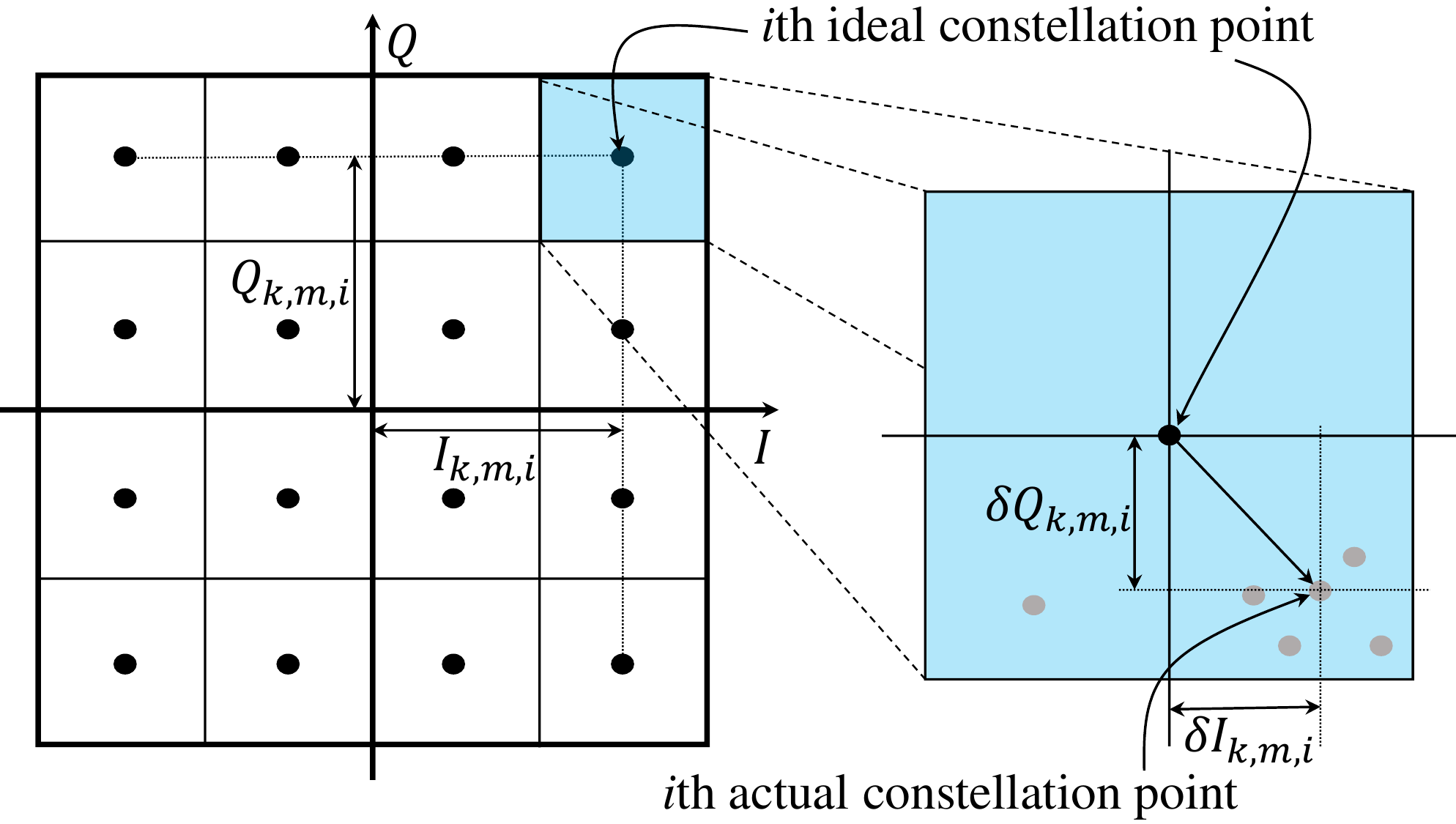}
	\vspace{0.2cm}
	\caption{Post-equalization SINR calculation from the constellation diagram.}
	\label{fig:SINR}
	\vspace{-0.5cm}
\end{figure}

The first component of the state space (i.e., the post-equalization SINR $\delta_{k,m}$) can be estimated from the constellation points in the data transmission phase. Let $\hat{\delta}_{k,m}$ be the estimated value of ${\delta}_{k,m}$. According to\cite{tr1012902001digital}, $\hat{\delta}_{k,m}$ can be calculated by
\begin{align}
\label{SINR_cal}
\hat{\delta}_{k,m}[n]=10\log_{10}\frac{\sum_{i=1}^{N^{\rm d}[n]}(I_{k,m,i}^2+Q_{k,m,i}^2)}{\sum_{i=1}^{N^{\rm d}[n]}(\delta I_{k,m,i}^2+\delta Q_{k,m,i}^2)},
\\[-1 cm]\nonumber
\end{align}
where $I_{k,m,i}$ and $Q_{k,m,i}$ are the ideal in-phase and quadrature components of the $i$th actual constellation point, respectively. On the other hand, $\delta I_{k,m,i}$ and $\delta Q_{k,m,i}$ are the in-phase and quadrature absolute differences between the $i$th actual and ideal constellation points, respectively. These terms are demonstrated in more detail in Fig. \ref{fig:SINR}. For each $i$th actual constellation point, the $\delta I_{k,m,i}$ and $\delta Q_{k,m,i}$ values are calculated by subtracting $I_{k,m,i}$ and $Q_{k,m,i}$ from the in-phase and quadrature components of the actual constellation point, respectively. Then, $\hat{\delta}_{k,m}$ can be calculated using Eq. (\ref{SINR_cal}). 
In fact, out of the two components of the state space, only the post-equalization SINR is observable at the UE as described above. The correlation coefficients, on the other hand, are controlled by the jammers (i.e., by Eq. (\ref{eqn:corr})), and are neither known nor directly observable by the UEs or the BS. The $\rho_{ij}^{\rm e}$ values can merely be indirectly observed by examining the largest $N_{\rm J}$ singular values of each covariance matrix ${\bf{R}}_{{\rm J}_{k}}^{\rm e}$ at the $K$ UEs. In general, small correlations between transmitted jamming signals result in relatively equal largest $N_{\rm J}$ singular values of ${\bf{R}}_{{\rm J}_{k}}^{\rm e}$, while large correlations result in massive differences between the largest $N_{\rm J}$ singular values.

Therefore, we formulate the problem as a partially observable SMDP (POSMDP)\cite{sutton2018reinforcement}, where the state in Eq. (\ref{eqn:State}) is replaced by the approximate state $\hat{\mathcal{S}}$ derived from the observations. Accordingly, we replace $s$ with $\hat{s}$ without further notice in all notations while keeping their meaning. For example, the optimal (approximate) state-action value function (introduced later) is denoted by $Q^*(\hat{s},a)$ instead of $Q^*(s,a)$. 

The observation space of the system is defined as:
\begin{align}
\label{eqn:Observation}
\mathcal{O}\delequal\big\{(\bar{\delta}_{k},\bar{\Lambda}_{l}):&\forall k\in (1,2,...,K),\nonumber\\
\forall m\in(1,2,...,M_{k}), &\forall l\in(1,2,...,N_{\rm J})
\big\},
	\\[-1 cm]\nonumber
\end{align}
where $\bar{\delta}_{k}=(1/M_{k})\sum_{m=1}^{M_{k}}\hat{\delta}_{k,m}$ is the average post-equalization SINR, $\Lambda_{k,l}$ is the $l$th largest singular value of ${\bf{R}}_{{\rm J}_{k}}^{\rm e}$, and  $\bar{\Lambda}_{l}=(1/K)\sum_{k=1}^{K}\Lambda_{k,l}$. To generate the observation at each epoch, $K$ average post-equalization SINR values $\bar{\delta}_{k}$ and $(K*N_{\rm J})$ singular values $\Lambda_{k,l}$ are calculated at the $K$ UEs. The observation is then obtained by concatenating $K$ average post-equalization SINR values $\bar{\delta}_{k}$ and $N_{\rm J}$ average singular value $\hat{\Lambda}_{l}$. Note that, similar to the state $\mathcal{S}$ mentioned above, the observation $\mathcal{O}$ is composed of $\bar{\delta}_{k}$ and $\bar{\Lambda}_{l}$ values in the most recent frame, such that
\begin{align}
o[n]\delequal\big[\bar{\delta}_{k}[n-1], \bar{\Lambda}_{l}[n-1]\big].
	\\[-1 cm]\nonumber
\end{align}
Similar to \cite{sutton2018reinforcement,mnih2015human}, we use the last $H$ observations and actions as the approximate state, i.e., $\hat{s}[n]\delequal\big[o[n], a[n-1],o[n-1],...,a[n-H]\big]$, where $H$ denotes the history length. This formalism, referred to as the $H$th-order history approach, generates a large but finite POSMDP, in which each sequence is a distinct approximate state. As a result, we can apply standard reinforcement learning techniques used in MDPs or SMDPs to find the optimal action given the current approximate state.
\subsubsection{Action}
At the end of each frame, an action is taken to determine which are the next $N^{\rm e}$ and $N^{\rm d}$ values, given the current approximate state. Let  ${\bf{\mathcal{N}}^{\rm e}}\delequal(N^{\rm e}_{1},N^{\rm e}_{2},...,N^{\rm e}_{L^{\rm e}})$ and ${\bf{\mathcal{N}}^{\rm d}}\delequal(N^{\rm d}_{1},N^{\rm d}_{2},...,N^{\rm d}_{L^{\rm d}})$ be the sets of $L^{\rm e}$ and $L^{\rm d}$ candidates for $N^{\rm e}$ and $N^{\rm d}$, respectively. The action space is defined as $\mathcal{A}\delequal\{a: a\in(1,2,...,L^{\rm e}\times L^{\rm d})\}$, and
\begin{align}
a = \begin{cases}
1, \qquad\qquad N^{\rm e}=N^{\rm e}_{1} \text{ and }N^{\rm d}=N^{\rm d}_{1},\\
2, \qquad\qquad N^{\rm e}=N^{\rm e}_{2} \text{ and }N^{\rm d}=N^{\rm d}_{1},\\
...\nonumber\\
L^{\rm e}\times L^{\rm d},  \quad N^{\rm e}=N^{\rm e}_{L^{\rm e}} \text{ and }N^{\rm d}=N^{\rm d}_{L^{\rm d}}.
\end{cases}
	\\[-1 cm]\nonumber
\end{align}
\subsubsection{Immediate Reward} The immediate reward is defined as the amount of data transmitted during the data transmission phase, and zero if one (or more) post-equalization SINR value during the data transmission phase is smaller than the minimum required post-equalization SINR. Specifically,
\begin{align}
\label{eqn:immediate reward}
r[n]=\begin{cases}
\sum_{k=1}^{K}\sum_{m=1}^{M_{k}}{N^{\rm d}[n]\log_{2}(1+{\delta}_{k,m}[n]}),\quad\text{when }{\delta}_{k,m}[n]\geq\delta_{\rm {min}},\nonumber\\
\qquad\qquad\quad\forall k\in (1,2,...,K),\forall m\in (1,2,...,M_{k}),\\
0,\qquad\qquad\rm{otherwise}.
\end{cases}
	\\[-1 cm]
\end{align}
\subsubsection{Optimization Formulation} Let $\pi:\hat{\mathcal{S}}\to{\mathcal{A}}$ denotes a policy which is a mapping function from the approximate states to the actions taken by the system. Our purpose is to find the optimal value of $\pi$, denoted by $\pi^*$, that maximizes the average long-term reward \cite{van2019optimal} of the BS-UE communication system. The optimization problem in Eq. (\ref{eqn:formulation_1}) is then converted into the optimization problem of $\pi^*$, expressed by
\begin{align}
	\label{eqn:formulation_2}
\max_{\pi}\quad\mathcal{R}(\pi)&=\lim_{N\to\infty}\frac{1}{N}\sum_{n=1}^{N}\mathbb{E}\{r[n]\}=\lim_{N\to\infty}\frac{1}{N}\sum_{n=1}^{N}\mathbb{E}\{r\{\hat{s}[n],\pi\{\hat{s}[n]\}\}\},
	\\[-1 cm]\nonumber
\end{align}
where $\mathcal{R}(\pi)$ denotes the average long-term reward of the system with the policy $\pi$.
\vspace{-0.5cm}
\subsection{Q-learning Technique}
\label{subsec:Q_learning}
This subsection introduces Q-learning \cite{watkins1992q}, a model-free reinforcement learning technique used to acquire the optimal policy $\pi^*$ without requiring prior information about jamming strategy as well as the channels condition.
Let $\mathcal{V}(s)$, $Q(s,a)$ and $\mathcal{G}({s},a)$ denote the state value function, the state-action value function, and the (state-dependent) action advantages function, respectively. The state value function $\mathcal{V}(s)$ is the expected cumulative reward of the system starting from the state $s$, illustrating ``how good'' it is for the system to be in the state $s$. The state-action value function $Q(s,a)$ demonstrates the expected discounted reward of the system in state $s$ selecting an action $a$. Finally, the action advantages function $\mathcal{G}({s},a)$ subtracts the state value function $\mathcal{V}(s)$ from the state-action value function $Q(s,a)$ to acquire the importance of each action.

The objective of the Q-learning is to find the optimal value of the state-action value function $Q(s,a)$, denoted by $Q^*(s,a)$, for all state-action pairs. Then, the optimal policy $\pi^*$ is obtained by $\pi^*=\argmax_{a}Q^*(s,a)$, meaning at each state, the system selects the action that maximizes the expected discounted reward. The optimal state-action value function $Q^*(s,a)$ can be achieved by executing an action, observing the next state and the immediate reward, and updating $Q(s,a)$ at each iteration. A detailed description of the Q-learning can be found in \cite{watkins1992q,sutton2018reinforcement}.

The Q-learning technique can be used to obtain the optimal policy $\pi^*$. However, this technique suffers from the slow-convergence problem. As mentioned in the previous subsection, the state $s$ is not fully observable by the BS and all the UEs. Therefore, we use the approximate state $\hat{s}$ to derive the optimal policy $\pi^*$. Nevertheless, the approximate state components (i.e., $\bar{\delta}_{k}$ and $\bar{\Lambda}_{l}$) are continuous values, resulting in an infinite dimension of the approximate state. Quantization of the approximate state components can reduce the dimension. However, a smaller quantization step size (i.e., for better accuracy) results in a larger approximate state space, causing the Q-learning technique to converge slowly. Moreover, the approximate state is composed of $H$ latest observations and actions, further increasing the dimension of the approximate state, and aggravating the slow-convergence issue. Therefore, in the following subsection, we adopt the deep dueling Q-learning technique\cite{wang2016dueling}, which uses a specialized neural network to efficiently obtain the optimal policy $\pi^*$.
\vspace{-0.5cm}
\subsection{Deep Dueling Q-learning Technique}
\label{Subsec:Deep Q_Learning}
This subsection presents the deep dueling Q-learning technique \cite{wang2016dueling} to deal with the partially observable and the slow-convergence problems of the Q-learning technique. In the Q-learning approach, $Q^*(s,a)$ is iteratively obtained and stored in a Q-table. However, in the deep dueling Q-learning technique, a neural network, referred to as the Q-network and denoted by $\mathcal{Q}$, is used as a nonlinear function approximator to estimate $Q^*(\hat{s},a)$. The input to the Q-network $\mathcal{Q}$ is the approximate state $\hat{s}$, and the output from $\mathcal{Q}$ is the optimal state-action value function $Q^*(\hat{s},a)$.

Let $\theta$ denotes the parameters of the Q-network $\mathcal{Q}$; the problem of finding $Q^*(\hat{s},a)$ becomes the problem of finding $\theta^*$, which are the optimal values of $\theta$. Accordingly, we include $\theta$ in the state value function, the state-action value function, and the action advantages function notations. For example, the state-action value function is now denoted by $Q(\hat{s},a;\theta)$, and its optimum is denoted by $Q^*(\hat{s},a;\theta^*)$. The deep dueling Q-learning technique to iteratively optimize $\theta$ is presented in Algorithm \ref{Alg:Deep Q_learning}. This algorithm is based on the one in \cite{mnih2015human}, and formed by the following techniques.
\begin{itemize}
	\item $\epsilon$-greedy action selection policy: At each training iteration, the system implements \textit{exploration} (i.e., by selecting a random action) with a probability of $\epsilon$, or \textit{exploitation} (i.e., by choosing the action that maximizes the current state-action value $Q(\hat{s},a;\theta)$) with a probability of $1-\epsilon$. The value of $\epsilon$ is large (e.g., $\epsilon=1$) at the starting iteration, and decays over the iterations as $\theta$ gets closer to the optimal value $\theta^*$.
	\item Experience replay: Instead of using instant transition at each iteration, the system stores the transitions in a memory pool $\mathbf{M}$ of size $M$ using the first-in-first-out (FIFO) protocol. At each training iteration, a random set is obtained from $\mathbf{M}$ to train the Q-network. This technique allows the previous transitions to be used more than once, hence improving the training data's efficiency. More importantly, by randomly selecting the training data from $\mathbf{M}$, the algorithm can remove the correlation between the consecutive training data.
	\item Target Q-network: This technique is performed by using a separate network for generating the target Q-values $y[j]$, as demonstrated by step \ref{step:q-network} in Algorithm \ref{Alg:Deep Q_learning}. The separate network with parameter $\hat{\theta}$ is named the target Q-network, and denoted by $\hat{\mathcal{Q}}$. Instead of updating at every iteration, the target Q-network $\hat{\mathcal{Q}}$ is only renewed every $C$ steps. As such, the primary Q-network is slowly updated, which helps to reduce the correlations between the estimated and target  Q-values, hence improving the stability of the deep dueling Q-learning technique.
	\item Mini-batch gradient descent\cite{goodfellow2016deep}: At each training iteration of the deep dueling Q-learning technique, instead of performing gradient descent using the whole data memory $\mathbf{M}$, the system randomly samples a mini-batch with $N_{\rm {mb}}$ samples from $\mathbf{M}$, and then performs mini-batch gradient descent on the mini-batch. By setting $N_{\rm {mb}}\ll D$, the training time can be reduced dramatically \cite{goodfellow2016deep}.
\end{itemize}
\begin{figure}[t]
	\centering
	\includegraphics[width=0.5\linewidth]{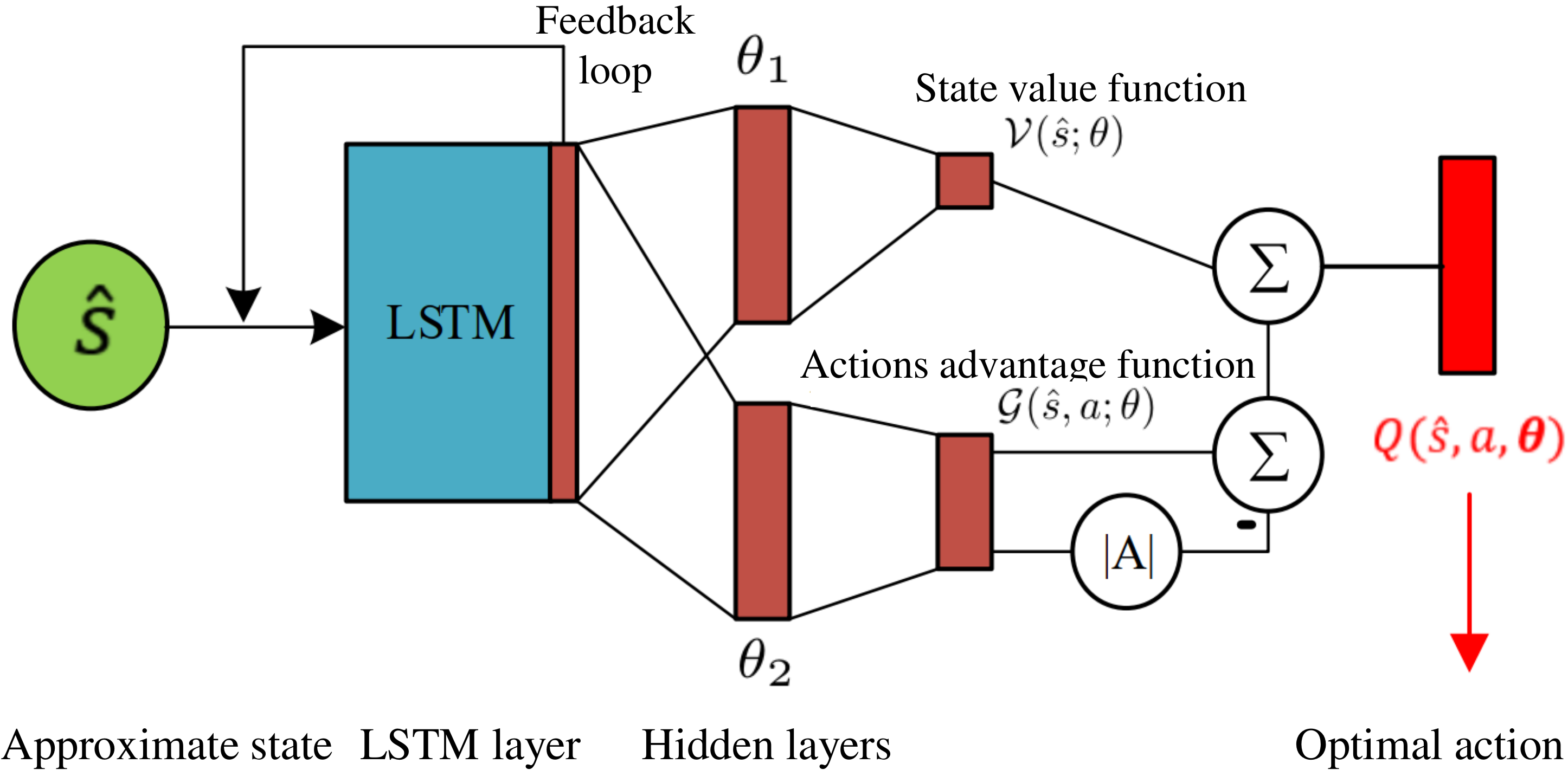}
	\vspace{0.2cm}
	\caption{Nullspace estimation and data transmission phases tunning \\using an LSTM-based deep dueling Q-network.}
	\label{fig:DL}
			\vspace{-0.5cm}
\end{figure}
\begin{algorithm}[t]
	\caption{Deep Dueling Q-learning Technique for Jamming Suppression.}
	\label{Alg:Deep Q_learning}
	\begin{algorithmic}[1]
		\State Initialize a memory $\mathbf{M}$ with capacity $M$. \label{step:1}
		\State Initialize $\mathcal{Q}$ and $\hat{\mathcal{Q}}$ with random weights $\theta$. \label{step:2}
		\For {iteration $i=1\ to\ I$}
		\State Select action \begin{align}
		a[i]= \begin{cases}
		\text{random action}, \quad\qquad\text{with probability }\epsilon\\
		\argmax_{a}Q(\hat{s}[i],a;\theta),\quad \text{otherwise}.
		\end{cases}
		\end{align}
		\State Perform $a[i]$, observe reward $r[i]$ and the next approximate state $\hat{s}[i+1]$.
		\State Store the transition $\{\hat{s}[i], a[i], r[i], \hat{s}[i+1]\}$ in $\mathbf{M}$.
		\State Randomly sample a mini-batch of $N_{\rm {mb}}$ transitions $\{s[j], a[j], r[j], s[j+1]\}$ from $\mathbf{M}$.
		\State Set
		\begin{align}
		Q(\hat{s}[j],a[j];\theta)=\mathcal{V}(\hat{s}[j];\theta)+\big(\mathcal{G}(\hat{s}[j],a[j];\theta)-\frac{1}{|\mathcal{A}|}\sum_{a[j]}^{}\mathcal{G}(\hat{s}[j],a[j];\theta)\big)
		\\[-1 cm]\nonumber
		\end{align} 
		\State Set $y[j]=r[j]+\gamma[i]\max_{a[j+1]} \hat{Q}(\hat{s}[j+1],a[j+1];\hat{\theta})$\label{step:q-network}
		\State Perform mini-batch gradient descent \cite{goodfellow2016deep} on $\{y[j]-Q(\hat{s}[j],a[j];\theta)\}^2$ with respect to $\theta$.
		\State Set $\hat{\mathcal{Q}}=\mathcal{Q}$ every $C$ iterations.
		\EndFor
	\end{algorithmic}
\end{algorithm}
\vspace{-0.5cm}
\subsection{Network Structure and Complexity Analysis}
In this subsection, we introduce the Long Short-Term Memory (LSTM) \cite{hochreiter1997long}-based deep dueling Q-network used in the deep dueling Q-learning technique. The network is illustrated in Fig. \ref{fig:DL}. Unlike conventional recurrent neural networks (RNN) that have difficulty learning the long-term dependencies of the inputs \cite{bengio1993problem}, the LSTM is capable of learning those dependencies, even with inputs consisting of more than 1000 discrete-time steps. That is because the LSTM is designed to avoid the ``vanishing gradient'' and exploding gradient'', which are the main problems in the training process of the RNN. Therefore, the LSTM is capable of solving sequential processing tasks not solvable by the RNN. On the other hand, the deep dueling network structure \cite{wang2016dueling} is developed to improve the convergence rate of the deep Q-learning technique, thanks to its innovative network structure. Specifically, the dueling structure contains two streams that separately estimate the state value and the advantages of actions. As such, the advantage streams of the network can concentrate on learning from only the relevant input (i.e., approximate state), hence improving the convergence rate. The input-output flow of the LSTM-based deep dueling Q-network in Fig. \ref{fig:DL} is as follows.

First, the approximate state $\hat{s}$ is fed into the LSTM layer as the input. The LSTM captures the time-varying characteristic of the correlations between transmitted jamming signals (i.e., by observing the average singular values $\bar{\Lambda}_{l}$) and the change in channels condition (i.e., by monitoring the average post-equalization SINR values $\bar{\delta}_{k}$). The output from the LSTM is then processed by two separated streams of fully-connected hidden layers to calculate the state value function $\mathcal{V}(\hat{s};\theta)$ and the advantages of actions $\mathcal{G}(\hat{s},a;\theta)$. The state-action value function $Q(\hat{s},a;\theta)$ is then calculated from $\mathcal{V}(\hat{s};\theta)$ and $\mathcal{G}(\hat{s},a;\theta)$ by \cite{wang2016dueling}
\begin{align}
Q(\hat{s},a;\theta)=\mathcal{V}(\hat{s};\theta)+\big(\mathcal{G}(\hat{s},a;\theta)-\frac{1}{|\mathcal{A}|}\sum_{a}^{}\mathcal{G}(\hat{s},a;\theta)\big),
\end{align}
where $|\mathcal{A}|$ denotes the dimension of the action space $\mathcal{A}$ (i.e., $|\mathcal{A}|=L^{\rm e}\times L^{\rm d}$).

The training process of the LSTM-based deep dueling Q-network has a computational complexity of $\mathcal{O}(W_{\theta})$, where $W_{\theta}$ denotes the total number of the Q-network's parameters; $W_{\theta}$ is given by\cite{sak2014long}
\begin{align}
\label{for:complexity}
W_{\theta}=&(4N_{\rm i}N_{\rm c}+4N_{\rm c}^2+N_{\rm c}N_{\rm {ol}}+3N_{\rm c})+(N_{\rm {ol}}N_{\theta_1}+N_{\rm {ol}}N_{\theta_2}+N_{\theta_2}|\mathcal{A}|+N_{\theta_1}),
	\\[-1 cm]\nonumber
\end{align}
where $N_{\rm i}$ is the number of input features, which is equal to the size of the $1$th-order history observation (i.e., $N_{\rm i}=K+N_{\rm J}+1$), $N_{\rm c}=H$ is the number of memory cells of the LSTM, $N_{\rm {ol}}$ is the output size of the LSTM layer, and $N_{\theta_1}$ and $N_{\theta_2}$ are the neuron number of the upper and lower separated fully-connected hidden layers in Fig. \ref{fig:DL}, respectively. The first line in Eq. (\ref{for:complexity}) shows LSTM's total number of parameters, while the total number of parameters from after the LSTM to the end of the LSTM-based deep dueling Q-network is given in the second line. Note that in Eq. (\ref{for:complexity}), for simplicity, the bias parameters in the neurons are ignored. Note also that the training process is performed by the BS, which has more computational power than the UEs. Moreover, when necessary, the training process can be offloaded to a cloud server connected to the BS through a backhaul link.

	\begin{table}[t]
	\small	
	\caption{\textsc{Parameters for the Deep Dueling Q-learning.}}
	\vspace{0.5 cm}
		{
		\centering
	\begin{tabular}{|P{5cm}|P{1.8cm}|P{6cm}|P{1.8cm}|}
		\hline 
		\textbf{Parameter} & \textbf{Value}&\textbf{Parameter}&\textbf{Value}\\ 
		\hline 
		LSTM's input feature size $N_{\rm i}$&$7$&LSTM's number of memory cell  $N_{\rm c}$ &6\\
		\hline 
		LSTM's output size $N_{\rm {ol}}$ &128&Fully connected layers size $N_{\theta_1}, N_{\theta_2}$ &(16, 16)\\ 
		\hline
		Mini-batch size $N_{\rm {mb}}$ &32&Memory size $M$ &10,000\\ 
		\hline
		Exploration rate $\epsilon$ range &[1.000, 0.1]&
		Exploration decay rate& 0.99\\ 
		\hline	
		Target network updating steps $C$&1000&Learning rate &0.01\\ 
		\hline
	\end{tabular}}
	\vspace{-0.5cm}
	\label{Tab:Q-learning param}
\end{table}
\vspace{-0.5cm}
\section{Simulation}
\label{Sec:Simulation Results}
\vspace{-0.5cm}
\subsection{Parameter Setting}
We consider a $200$ kHz bandwidth (i.e., corresponding to a symbol duration of $5$ $\mu\rm{s}$) MIMO system containing a BS and $K=4$ UEs, each receiving $M_{k}=3$ signal streams from the BS. The BS and all the UEs have ULA array structures with $12$ and $8$ antennas, respectively (i.e., $N_{\rm T}=12$ and $N_k=8, \forall k\in (1,2,3,4)$). The carrier frequency is $447$ MHz and the transmitted power is $P_{\rm T}=44$ dBm. The received signals at the receivers are sampled at a sampling rate of $400$ kHz. The legitimate signals are modulated using the $16$-Quadrature Amplitude Modulation (QAM) technique. To evaluate the system's performance more directly, we do not use any forward error correction (FEC) coding. As presented in \cite{sousa1994delay}, the standard root mean square (RMS) delay spread in the urban, suburban, combined, and rural areas are all equal or smaller than $0.73$ $\mu\rm{s}$. Therefore, the ratio between the symbol duration and the RMS delay spread is $5/0.73\approx7$. As a result, the intersymbol interference (ISI) impact on BS-UE signal decoding is negligible. For small-scale fading, as presented in Section \ref{Sec:System_model}, a flat fast fading Rayleigh model in \cite{xiao2006novel} is used to simulate the jammers-UEs and BS-UEs channels. We assume the jammers-UE and BS-UE relative velocities are around $20$ km/h, corresponding to a Doppler frequency of $F_{\rm d}=8.28$ Hz. The number of propagation paths for each BS-UE link is $N^{\rm {p}}_{k}=8, \forall k\in(1,2,3,4)$. For the large-scale path-losses $\eta_{k}$ and $\eta_{k,j}$, we use the COST 231 Hata model \cite[Ch.~4]{damosso1999digital}, and assume $h_{\rm {T}}=50$ m, $h_{\rm J}=h_{\rm {UE}}=2$ m, and $d_{\rm J}=d_{\rm {T}}=100$ m, where $h_{\rm {T}},h_{\rm J},h_{\rm {UE}},d_{\rm J}$, and $d_{\rm {T}}$ denote BS's height, jammers' height, UEs' height, the initial distance from each jammer to the UEs, and initial BS-UE distance, respectively. Without loss of generality, we assume the BS and the jammers are stationary, and the heights of all the UEs are unchanged while moving.

There are $N_{\rm J}=2$ jammers with intentionally time-varying correlations between their transmitted signals. As presented, the time-varying correlations between the transmitted jamming signals cause a ``virtual change'' in the jamming channel. For demonstration, the correlation $\rho_{12}$ between the two jammers is assumed to repeatedly and linearly decrease from $1$ to $0.8$ over $5000$ samples. Specifically, the correlation at $p$th sample is given by
\begin{align}
\label{eqn:jam_cor}
 \rho_{12}(p)=\mathcal{J}(p)=1-0.2(p-5000\times \floor {p/5000}),
\end{align}
where $\floor{.}$ denotes the floor function, determining the closest smaller integer.
 
The candidate sets for $N^{\rm e}$ and $N^{\rm d}$ are ${\bf{\mathcal{N}}^{\rm e}}=(10,20,30,40)$ and ${\bf{\mathcal{N}}^{\rm d}}=(200,250,300,350)$ samples (i.e., corresponding to $(100,125,150,175)$ symbols per each data transmission phase), respectively. The selections for $N^{\rm e}$ and $N^{\rm d}$ follow the ETSI standard for Terrestrial Trunked Radio (TETRA)\cite{etsi1997terrestrial}. The parameters of the LSTM-based deep dueling Q-network and its training parameters are given in Table. \ref{Tab:Q-learning param}, which are similar to those in \cite{van2019jam,mnih2015human}. 

To illustrate the advantage of the proposed technique, we compare the following schemes: 
\begin{itemize}
	\item \textit{Upper bound}: The system is assumed to perfectly nullify the jamming signals by using the estimated beam-forming matrix. For illustration purposes, instead of being calculated using the received jamming signal in the nullspace estimation phase, the beam-forming matrix is directly estimated using the jamming signal in the data transmission phase (which is, in reality, unknown to the system). 
	\item \textit{Fixed action}: The system uses a fixed pair of values for $N^{\rm e}$ and $N^{\rm d}$. The performance metrics are calculated by averaging the performance of ($L^{\rm e}\times L^{\rm d}$) action choices;
	\item \textit{Heuristic approach \cite{hoang2021suppression}}: The system uses the jamming nullification technique in \cite{hoang2021suppression}, in which the residual jamming signals are measured, and the estimated beam-forming matrix is updated whenever the residual exceeds a predefined value;
	\item\textit{Proposed technique}: The values of $N^{\rm e}$ and $N^{\rm d}$ are determined by the optimal policy obtained using the proposed deep dueling Q-learning technique. 
\end{itemize}
\vspace{-0.5cm}
\subsection{Simulation Result}
For a fair comparison, we average the effective spectral efficiency and the outage probability over $N=5000$ frames and the $K$ UEs, and denote them by $C_{\rm {av}}^{\rm {eff}}$ and $p_{\rm {av}}^{\rm ot}$, respectively. We have
	\begin{align}
	C_{\rm {av}}^{\rm {eff}}&=\frac{1}{NKM_{k}}\sum_{n=1}^{N}\sum_{k=1}^{K}\sum_{m=1}^{M_{k}}C_{k,m}^{\rm eff}[n],\quad p_{\rm {av}}^{\rm ot}=\frac{1}{NKM_{k}}\sum_{n=1}^{N}\sum_{k=1}^{K}\sum_{i=1}^{M_{k}}\mathbf{1}_{\delta_{k,m}[n]<\delta_{\rm {min}}},\nonumber\\
	\text{where }& \mathbf{1}_{\delta_{k,m}[n]<\delta_{\rm {min}}}=\begin{cases}
	1,\qquad\qquad\delta_{k,m}[n]<\delta_{\rm {min}},\\\nonumber
	0,\qquad\qquad \text{otherwise},
	\end{cases}
	\end{align}
	and $\delta_{\rm {min}}=11.8$ dB is the minimum required SINR.
\begin{figure}[t]
	\centering
	\includegraphics[width=0.45\linewidth]{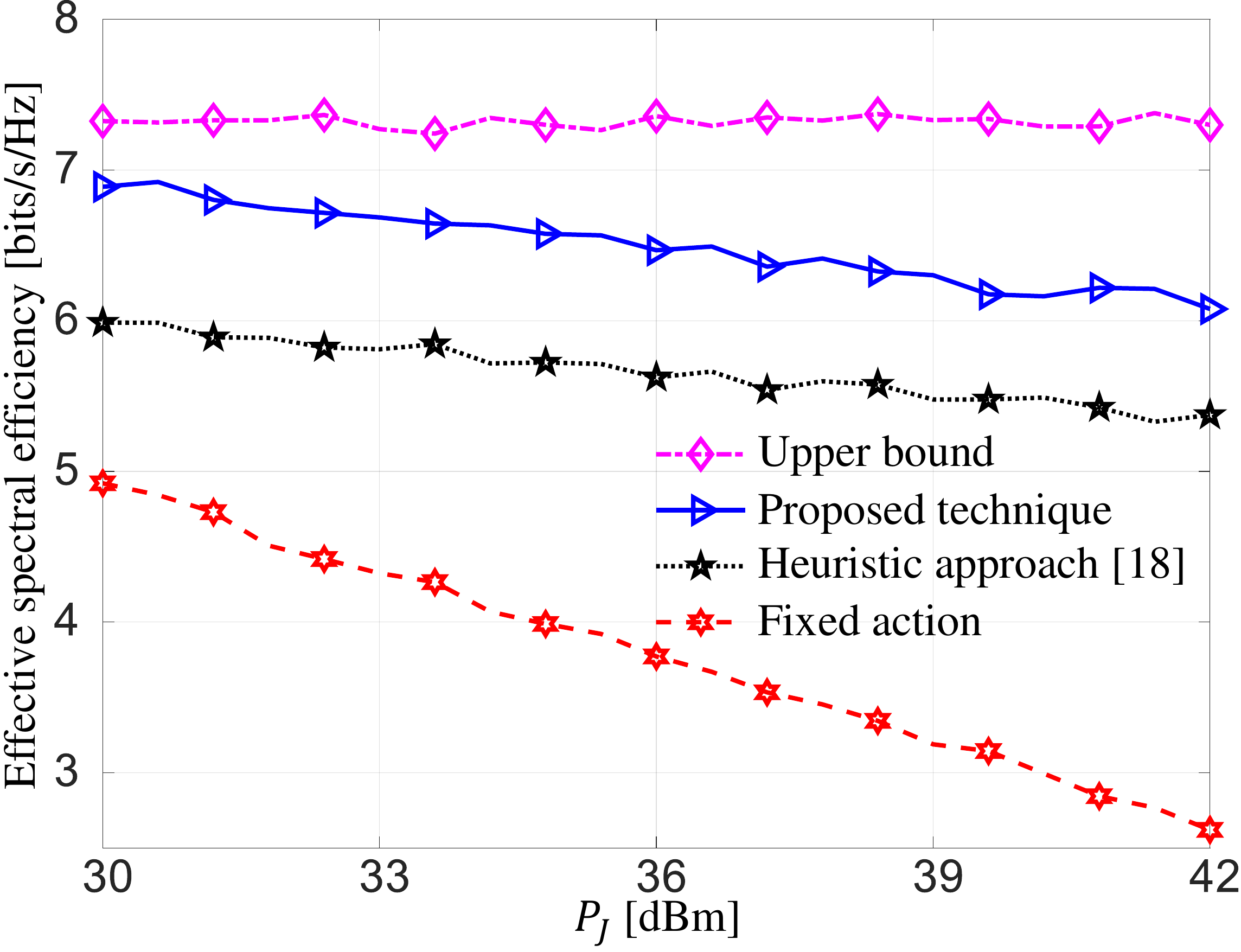}
	\vspace{0.2cm}
	\caption{Effective spectral efficiency for different techniques and jamming powers.}
	\label{fig:spectral efficiency}
		\vspace{-0.5cm}
\end{figure}
\begin{figure}[t]
	\centering
	\includegraphics[width=0.45\linewidth]{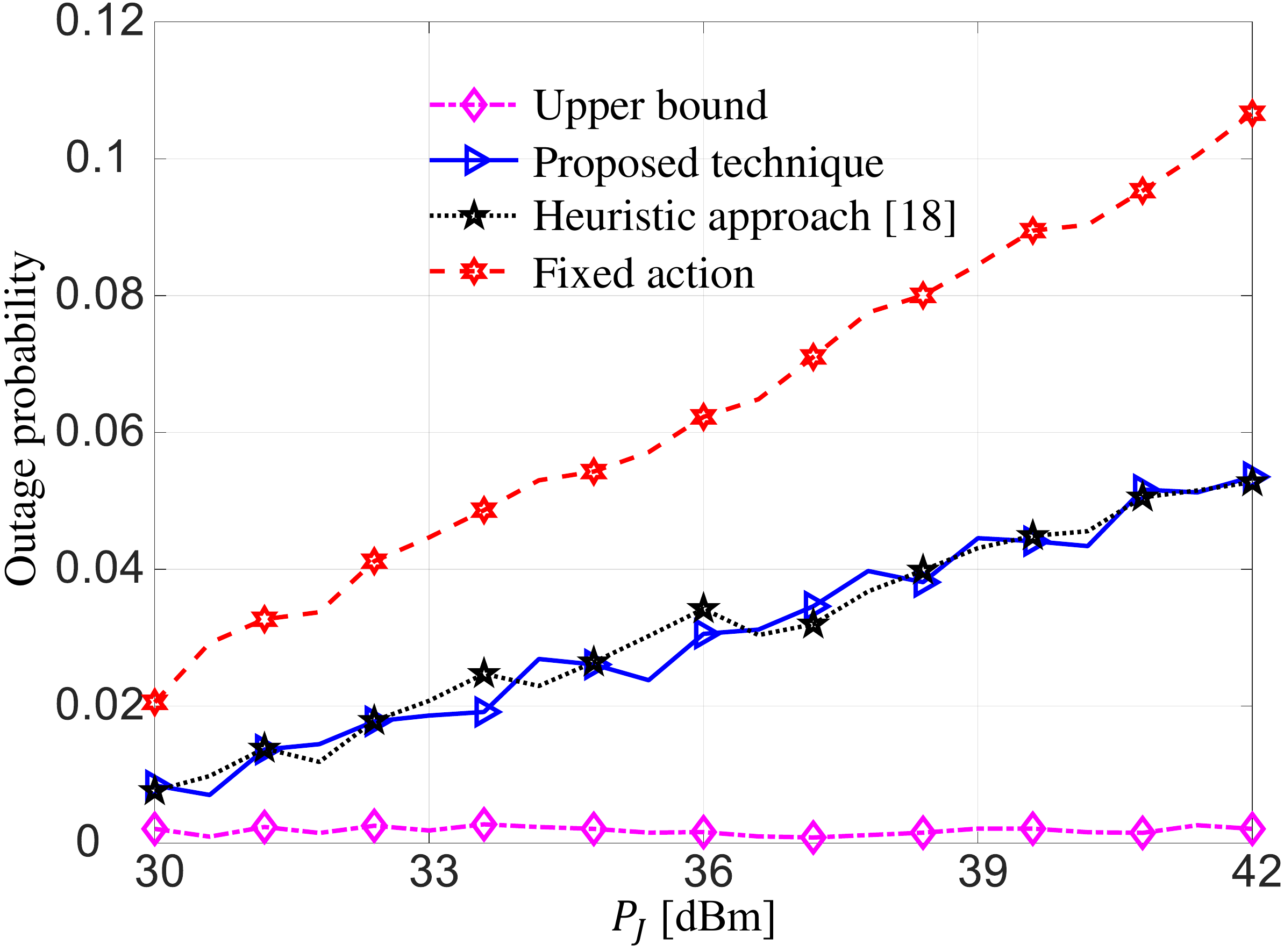}
	\vspace{0.2cm}
	\caption{Outage probability for different techniques and jamming powers.}
	\label{fig:outage}
		\vspace{-0.5cm}
\end{figure}
\subsubsection{Effective Spectral Efficiency Analysis}
Fig. \ref{fig:spectral efficiency} shows the average effective spectral efficiency $C_{\rm {av}}^{\rm {eff}}$ of each BS-UE communication link for different jamming nullification techniques and different jamming power $P_{\rm {J}}$. Note that the jamming power is calculated by the average variance of all the jammers (i.e., $P_{\rm {J}}=(1/N_{\rm J})\sum_{j=1}^{N_{\rm J}}\sigma_{{\rm J}_{j}}^2$). As can be seen, the proposed deep dueling Q-learning technique achieves the highest average effective spectral efficiency for all values of the jamming power, thanks to its ability to efficiently adjust the $N^{\rm e}$ and $N^{\rm d}$ values according to the change in the correlations and channel conditions. On the other hand, the other two techniques have several limitations. The technique in \cite{hoang2021suppression} spends an excessive amount of time monitoring the residual jamming signals and estimating the beam-forming matrix, thus reducing the data transmission time. Meanwhile, the \textit{fixed action} technique cannot adapt to the change in the channel conditions, and more importantly, that in the correlations between transmitted jamming signals, resulting in an ineffective estimated beam-forming matrix ${\hat{\bf{F}}}_{k}$. These limitations of \cite{hoang2021suppression} and \textit{fixed action} result in lower spectral efficiencies of the communication system. Note that there is a gap between the spectral efficiency of the proposed technique and that of the \textit{upper bound}. This is because, in the \textit{upper bound} case, the jamming signal is completely canceled out without needing to adjust the values of $N^{\rm e}$ and $N^{\rm d}$. Meanwhile, for the proposed technique, $N^{\rm e}$ may have to be increased to improve the estimation accuracy of ${\hat{\bf{F}}}_{k}$, hence reducing the effective spectral efficiency.
\subsubsection{Outage Probability Analysis}
Fig. \ref{fig:outage} illustrates the outage probability of the systems using the three mentioned techniques and different values of the jamming power. As can be seen, the proposed deep dueling Q-learning technique and the techniques in \cite{hoang2021suppression} have very similar outage probabilities, and are much lower than that of the \textit{fixed action} technique. This is because the \textit{fixed action} technique cannot adapt to the change of the correlations between transmitted jamming signals and channels condition, resulting in many outage frames because of excessive jamming residuals. On the other hand, both \textit{heuristic approach} and the proposed technique effectively nullify the jamming signals. However, as mentioned above, the \textit{heuristic approach} in \cite{hoang2021suppression} spends an excessive amount of time monitoring the residual jamming signals and estimating the beam-forming matrix, resulting in a much lower spectral efficiency. Note that there is a difference in the outage probability between the proposed technique and the \textit{upper bound}, because the jamming signal, in fact, cannot be entirely canceled out as in the \textit{upper bound} case. 
\begin{figure}[t]
	\centering
	\includegraphics[width=0.45\linewidth]{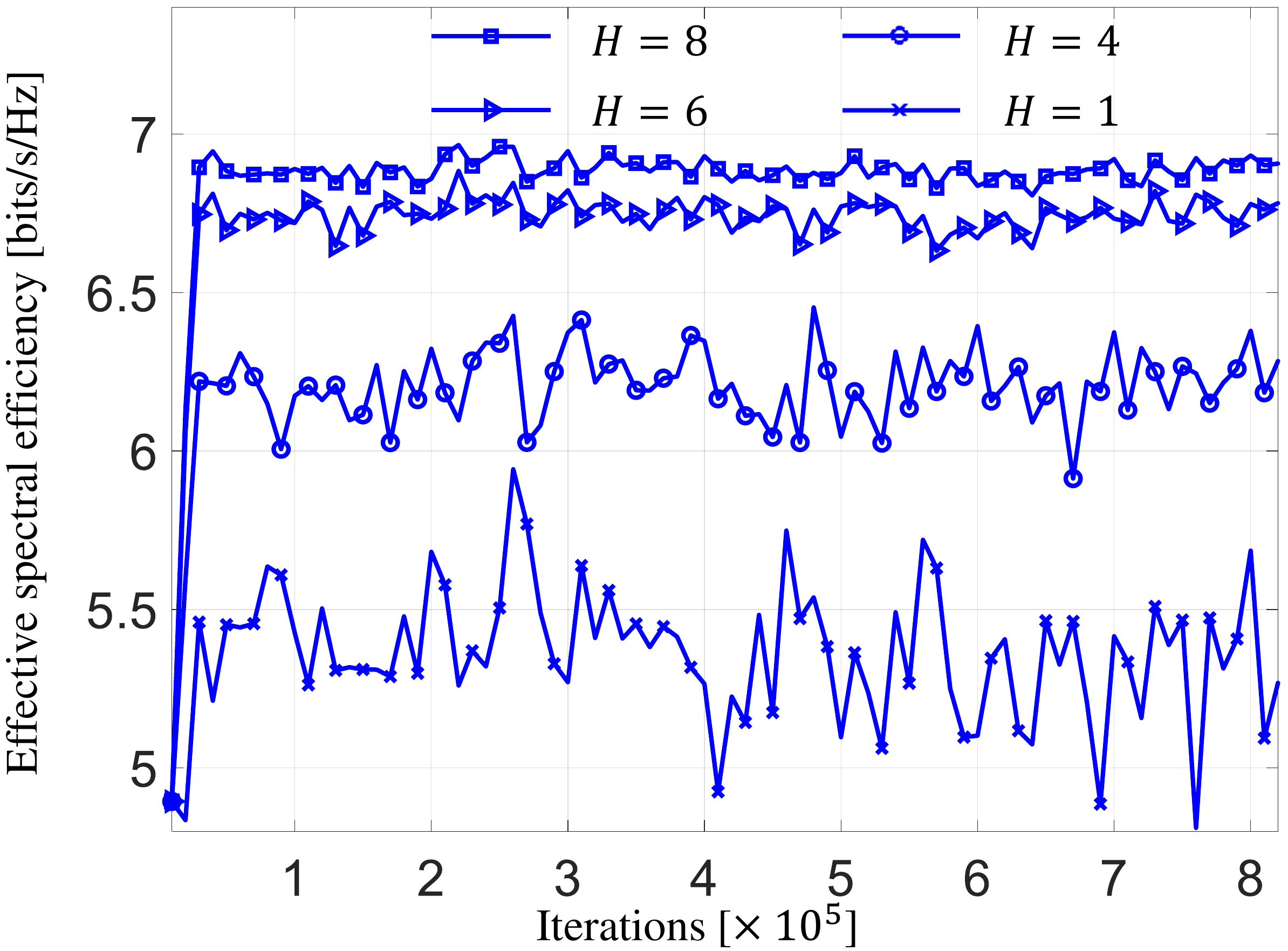}
	\vspace{0.2cm}
	\caption{Spectral efficiency convergence rate for different history lengths.}
	\label{fig:convergent_spectral efficiency}
			\vspace{-0.5cm}
\end{figure}
\begin{figure}[t]
	\centering
	\includegraphics[width=0.45\linewidth]{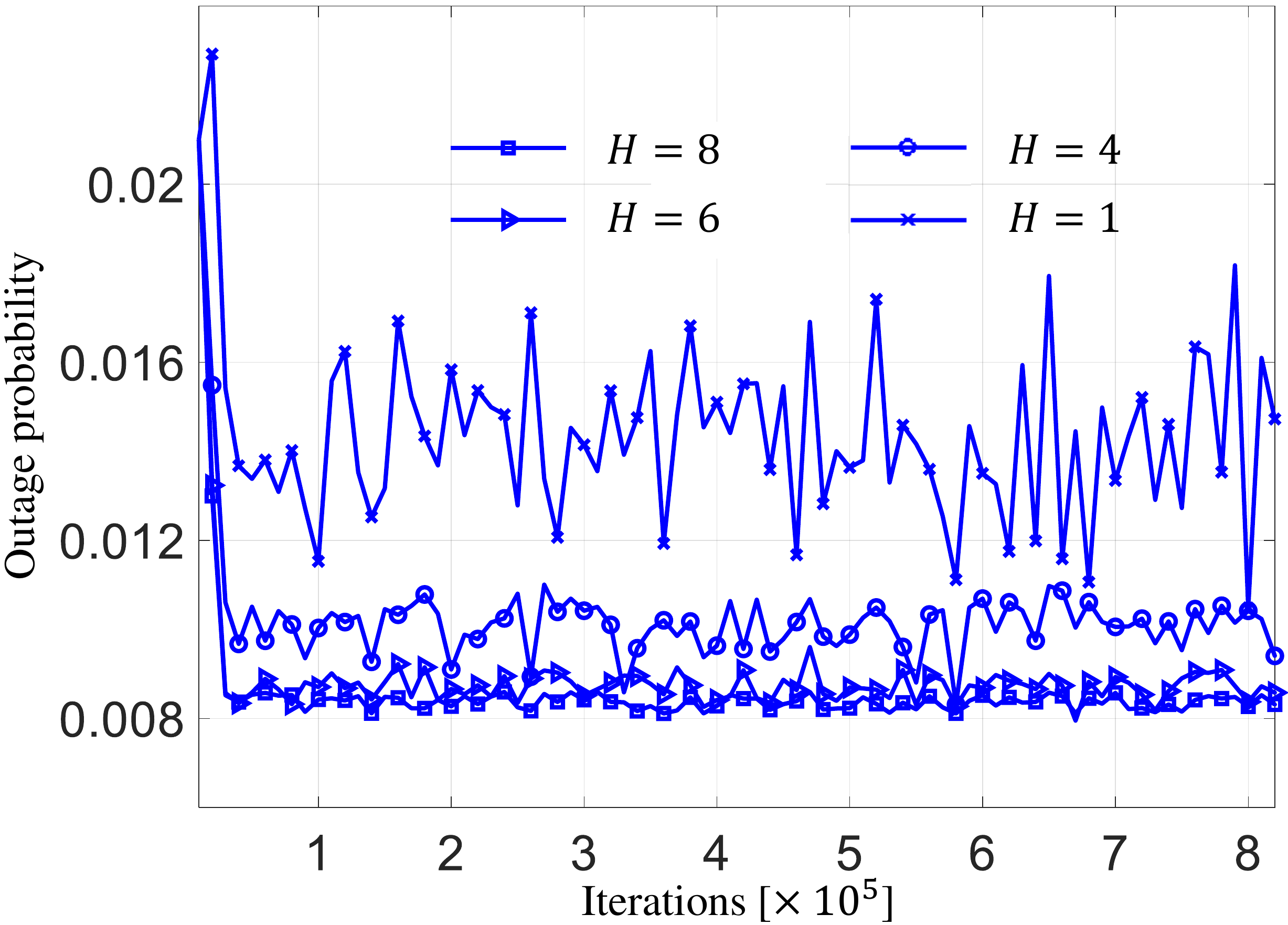}
	\vspace{0.2cm}
	\caption{Outage probability convergence rate for different history lengths.}
	\label{fig:convergent_outage}
		\vspace{-0.5cm}
\end{figure}
\subsubsection{Impact of History Length $H$}
Fig. \ref{fig:convergent_spectral efficiency} and Fig. \ref{fig:convergent_outage} illustrate the impact of the historical length $H$ on the convergence of the proposed deep dueling Q-learning technique. The jamming power used to generate these figures is $30$ dBm. As can be seen, the deep dueling Q-learning technique converges after around $2\times 10^4$ iterations. Note that due to the randomness of the environment and the initialization of the LSTM-based deep neural network, the number of iterations required for convergence may vary. Note also that, at the converged status, a longer history length $H$ results in a higher effective spectral efficiency and a lower outage probability. However, increasing the value of $H$ also increases the computational complexity of the deep dueling Q-learning technique, as demonstrated by Eq. (\ref{for:complexity}). Another interesting point is that, the spectral efficiency and the outage probability do not dramatically improve when $H$ increases from $6$ to $8$. Therefore, using $H=6$ can balance the technique's performance and computational complexity.
\subsubsection{Adaptability to the Change in Jamming Strategy}
\begin{figure}[t]
	\centering
	\includegraphics[width=0.45\linewidth]{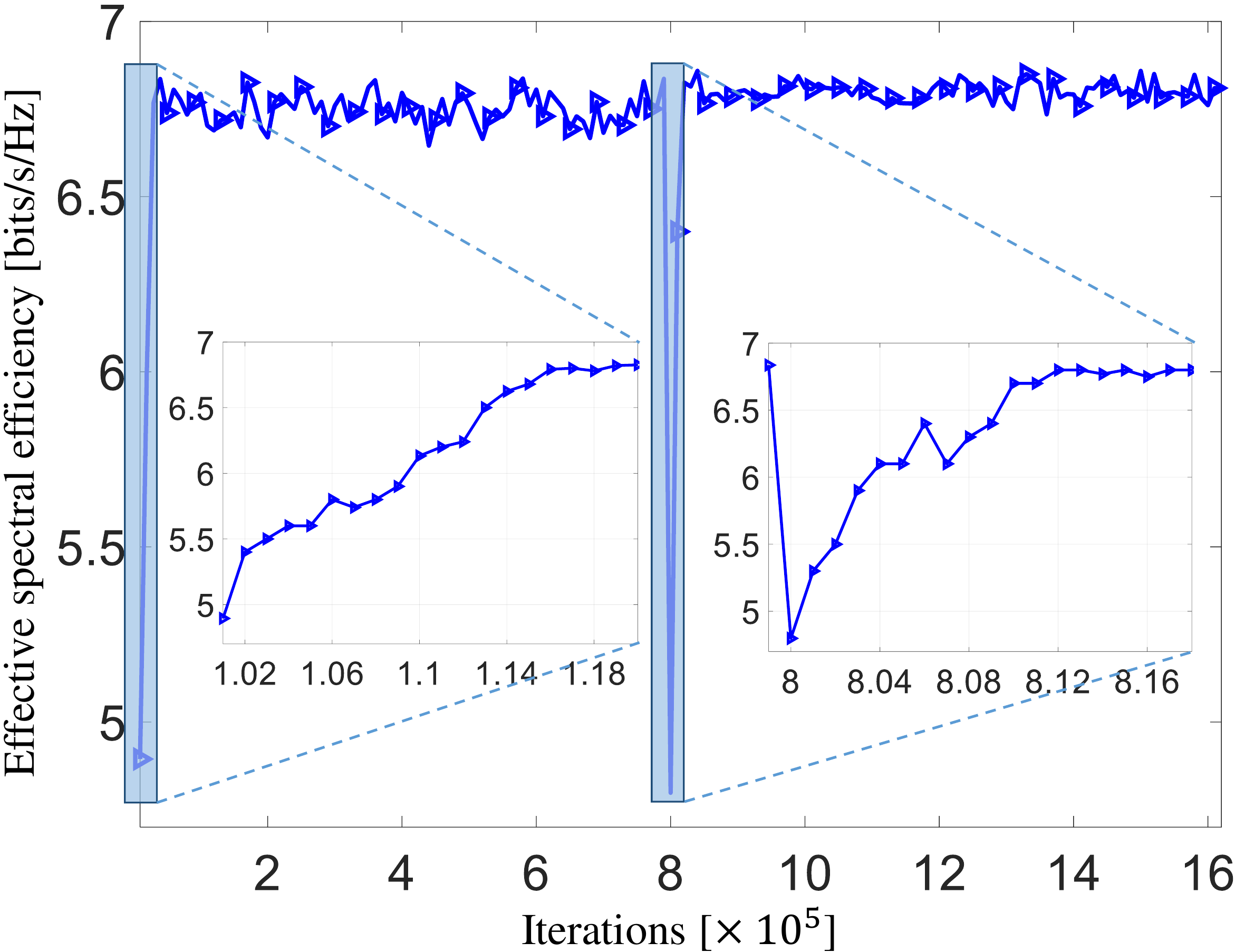}
	\vspace{0.2cm}
	\caption{Adaptability to the change in the jamming strategy.}
	\label{fig:strategy_spectral}
	\vspace{-0.6cm}
\end{figure}

\begin{figure}[t]
	\centering
	\includegraphics[width=0.45\linewidth]{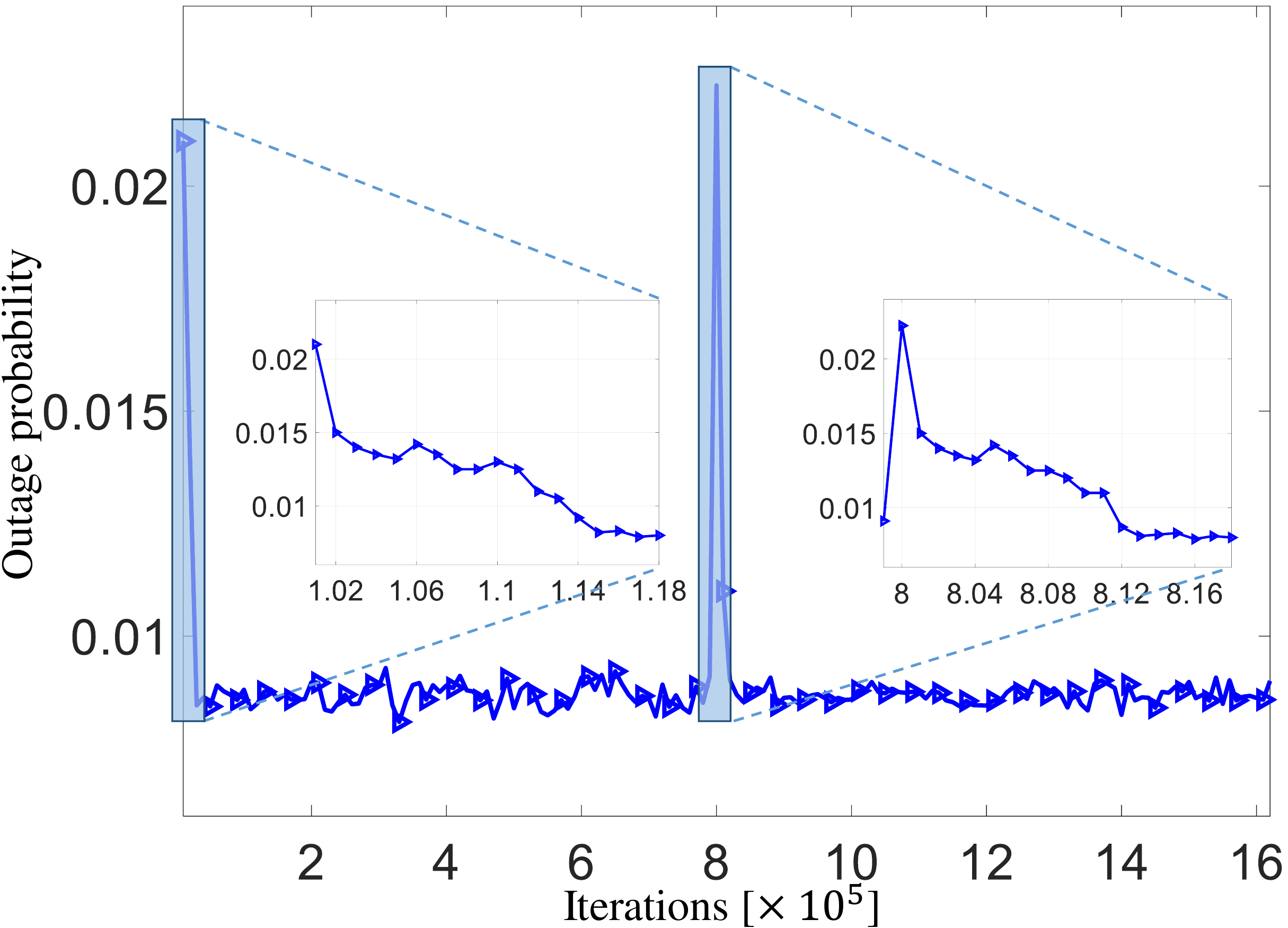}
	\vspace{0.2cm}
	\caption{Adaptability to the change in the jamming strategy.}
	\label{fig:strategy_outage}
		\vspace{-0.6cm}
\end{figure}

Fig. \ref{fig:strategy_spectral} and Fig. \ref{fig:strategy_outage} illustrate the capability of the deep dueling Q-learning technique to adapt the optimal anti-jamming policy when the jamming strategy is changed. Specifically, from the $(8\times10^5)$th iterations, instead of linearly decreasing as in Eq. (\ref{eqn:jam_cor}), the correlation is linearly increase from $0.8$ to $1$ as
\begin{align}
\rho_{12}(p)=\mathcal{J'}(p)=0.8+0.2[(p-8\times 10^5)-5000\floor {(p-8\times 10^5)/5000}].\nonumber
\end{align}
As can be seen, the deep dueling Q-learning technique can adapt to the change in the jamming strategy by quickly re-obtain the convergence for both effective spectral efficiency and outage probability. In particular, the re-establishments of the convergence status (i.e., from the $(8\times 10^5)$th iteration where the jamming strategy changes) are even slightly faster than the first convergence (i.e., from the first iteration). This is because the Q-network at the $(8\times 10^5)$th iteration is ``initialized'' by the network parameters obtained from the previous training process. Therefore, this ``initialization'' performs better than the random initialization at the first iteration when the training process begins. 
\vspace{-0.5cm}
\section{Conclusions}
\label{Sec:conclusion}
We have examined the impact of time-varying correlations between transmitted jamming signals on jamming nullification. We have demonstrated that using an incorrectly estimated beam-forming matrix can waste the receiver's degree-of-freedoms without achieving effective jamming suppression. We have proposed the deep dueling Q-learning technique to effectively estimate the beam-forming matrix and nullify the jamming signals designed to aggravate the jamming impact. The dueling network architecture allows our proposed technique to quickly obtain the optimal solution against the jammers, therefore very effective in dealing with unknown and time-varying jamming strategies. The simulation results demonstrate that our proposed deep dueling Q-learning technique achieves a higher effective spectral efficiency and a lower outage probability than the existing techniques.
\appendices
\vspace{-0.5cm}
\section {Proof of Theorem $2$}
\label{appen_theorem2}
Assuming the BS-UE channel can be estimated correctly, the received signal at the $k$th UE after zero-forcing equalization can be expressed by
\begin{align}
\label{eqn:received_signal_ZF}
{\bf{y}}_{k}^{\rm ZF}&=(\tilde{{\bf{H}}}_k^H\tilde{{\bf{H}}}_k)^{-1}\tilde{{\bf{H}}}_k^H(\sqrt{P_{\rm T}}\tilde{{\bf{H}}}_k{\bf{x}}_{k}+\tilde{\bf{Z}}_{k}{\bf{x}}_{\rm J}+{\tilde{\bf{w}}})=\sqrt{P_{\rm T}}{\bf{x}}_{k}+{\bf{A}}_k^{\rm {ZF}}(\tilde{\bf{Z}}_{k}{\bf{x}}_{\rm J}+{\tilde{\bf{w}}}),
\end{align}
where ${{\bf{A}}_k^{\rm {ZF}}}=(\tilde{{\bf{H}}}_k^H\tilde{{\bf{H}}}_k)^{-1}\tilde{{\bf{H}}}_k^H$ is the zero-forcing equalizer, $\tilde{{\bf{H}}}_k$, $\tilde{{\bf{Z}}}_k$, and ${\tilde{\bf{w}}}$ are the BS-$k$th UE equivalent channel, jammer-$k$th UE equivalent channel, and equivalent noise, respectively. The values of $\tilde{{\bf{H}}}_k$, $\tilde{{\bf{Z}}}_k$, and ${\tilde{\bf{w}}}$ can be expressed by
\begin{align}
&\tilde{{\bf{H}}}_k=\tilde{{\bf{H}}}_k^{\rm {wbf}}={{\bf{H}}}_k{{\bf{P}}}_k,\quad \tilde{{\bf{Z}}}_k={{\bf{Z}}}_k,\quad{\tilde{\bf{w}}}={{\bf{w}}},\qquad\quad\qquad\text{without beam-forming},\nonumber\\
&\tilde{{\bf{H}}}_k=\tilde{{\bf{H}}}_k^{\rm {bf}}={\hat{\bf{F}}}_{k}{\bf{H}}_{k}{\bf{P}}_{k},\quad \tilde{{\bf{Z}}}_k={\hat{\bf{F}}}_{k}{{\bf{Z}}}_k,\quad{\tilde{\bf{w}}}={\hat{\bf{F}}}_{k}{\bf{w}},\qquad\:\text{with beam-forming},\nonumber
\end{align}

First, considering the case without beam-forming, from Eq. (\ref{eqn:received_signal_ZF}), the post-equalization SINR of the $m$th stream for the $k$th UE can be expressed by
	\begin{align}
	\label{snr1}
	\delta_{k,m}^{{\rm {wbf}}}=\frac{{P_{\rm T}}{\mathbb{E}}({{x}}^2_{k,m})}{{\mbox{Var}}\{[{{\bf{A}}_k^{\rm {ZF}}}({\bf{Z}}_{k}{\bf{x}}_{\rm J}+{\bf{w}})]_m\}}=\frac{{P_{\rm T}}}{{\mbox{Var}}\{[{{\bf{A}}_k^{\rm {ZF}}}({\bf{Z}}_{k}{\bf{x}}_{\rm J}+{\bf{w}})]_m\}}.\qquad\;	
	\\[-0.7 cm]\nonumber
	\end{align}
Since ${{\bf{A}}_k^{\rm {ZF}}}$ and $({\bf{Z}}_{k}{\bf{x}}_{\rm J}+{\bf{w}})$ are independent, we evaluate their covariance separately. Specifically,
\begin{align}
\label{cov1}
{\mbox{Cov}}({{\bf{A}}_k^{\rm {ZF}}})&={\mathbb{E}}[(\tilde{{\bf{H}}}_k^{{\rm {wbf}}^H}\tilde{{\bf{H}}}_k^{{\rm {wbf}}})^{-1}\tilde{{\bf{H}}}_k^{{\rm {wbf}}^H}\tilde{{\bf{H}}}_k^{{\rm {wbf}}}(\tilde{{\bf{H}}}_k^{{\rm {wbf}}^H}\tilde{{\bf{H}}}_k^{{\rm {wbf}}^H})^{-1}]\qquad\qquad\qquad\qquad\nonumber\\
&={\mathbb{E}}[(\tilde{{\bf{H}}}_k^{{\rm {wbf}}^H}\tilde{{\bf{H}}}_k^{{\rm {wbf}}^H})^{-1}]\overset{\mathrm{(a)}}{=}
\frac{\eta_{k}{\bf{I}}_{M_{k}}}{N_k-M_{k}}.
\end{align}
where (a) follows because $\tilde{{\bf{H}}}_k^{{\rm wbf}}\in \mathbb{C}^{N_k \times M_{k}}\sim\mathcal{CN}({\mathbf{0}},1/\sqrt{\eta_{k}})$ and hence its sample covariance $\tilde{{\bf{H}}}_k^{{\rm wbf}^H}\tilde{{\bf{H}}}_k^{{\rm wbf}}$ has a complex Whishart distribution with ${\mathbb{E}}[(\tilde{{\bf{H}}}_k^{{\rm wbf}^H}\tilde{{\bf{H}}}_k^{{\rm wbf}})^{-1}]=\eta_{k}{\bf{I}}_{M_{k}}/(N_k-M_{k})$\cite{lozano2003multiple,wang2007performance}. On the other hand, the covariance of $({\bf{Z}}_{k}{\bf{x}}_{\rm J}+{\bf{w}})$ is the total power of the noise and the jamming signals at the UE receiver, i.e.,
\begin{align}
\label{cov2}
{\mbox{Cov}}({\bf{Z}}_{k}{\bf{x}}_{\rm J}+{\bf{w}})=\big(\sigma_w^2+\sum_{j=1}^{N_J}{\eta_{k,j}}\sigma_{{\rm J}_{j}}^2\big){\bf{I}}_{N_k}.\qquad\qquad\,\:\qquad\qquad\qquad\quad
\end{align}
Substitute Eq. (\ref{cov1}) and Eq. (\ref{cov2}) into Eq. (\ref{snr1}) yields
	\begin{align}
\label{snr2}
\delta_{k,m}^{{\rm {wbf}}}=\frac{{P_{\rm T}(N_k-M_{k})}}{\eta_{k}(\sigma_w^2+\sum_{j=1}^{N_J}{\eta_{k,j}}\sigma_{{\rm J}_{j}}^2)}.\qquad\qquad\qquad\qquad\qquad\quad
\\[-1 cm]\nonumber
\end{align}
The spectral efficiency of the $m$th stream for the $k$th UE is therefore expressed by
\begin{align}
\label{spectral_wbf}
	C_{k,m}^{\rm wbf}=&\log_{2}\big[1+\frac{{P_{\rm T}(N_k-M_{k})}}{\eta_{k}(\sigma_w^2+\sum_{j=1}^{N_J}{\eta_{k,j}}\sigma_{{\rm J}_{j}}^2)}\big].\qquad\qquad\qquad\quad
		\\[-1 cm]\nonumber
\end{align}

When the estimated beam-forming matrix is applied, the post-equalization SINR of the $m$th stream for the $k$th UE is
\begin{align}
\delta_{k,m}&=\frac{P_{\rm T}}{{\mbox{Var}}\{[{{\bf{A}}_k^{\rm {ZF}}}{\hat{\bf{F}}}_{k}({\bf{Z}}_{k}{\bf{x}}_{\rm J}+{\bf{w}})]_m\}},\label{snr_lb_1}\\
&=\frac{P_{\rm T}}{{\mbox{Var}}\{[{{\bf{A}}_k^{\rm {ZF}}}(\delta{\bf{F}}_{k}{\bf{Z}}_{k}{\bf{x}}_{\rm J}+{\hat{\bf{F}}}_{k}{\bf{w}})]_m\}}.\qquad\qquad\qquad\quad\label{snr_lb}
\\[-1 cm]\nonumber
\end{align}
where $\delta{\bf{F}}_{k}={\hat{\bf{F}}}_{k}-{\bf{F}}_{k}$ is the estimation error of ${\bf{F}}_{k}$.

When the estimated beam-forming matrix is estimated perfectly (i.e., $\delta{\bf{F}}_{k}=\bf{0}$), the jamming signals are cancelled totally, Eq. (\ref{snr_lb}) becomes
	\begin{align}
\label{snr3}
\delta_{k,m}^{{\rm {ub}}}&=\frac{{P_{\rm T}}}{{\mbox{Var}}[{{(\bf{A}}_k^{\rm {ZF}}}{\hat{\bf{F}}}_{k}{\bf{w}})_m]}.\quad\qquad\qquad\qquad\qquad\qquad\quad\,
\\[-1 cm]\nonumber
\end{align}
Similar to the case without the estimated beam-forming matrix, we evaluate the covariance of ${{\bf{A}}_k^{\rm {ZF}}}{\hat{\bf{F}}}_{k}$ and ${\bf{w}}$ in Eq. (\ref{snr3}) separately. Specifically,
\begin{align}
\label{cov5}
{\mbox{Cov}}({{\bf{A}}_k^{\rm {ZF}}}{\hat{\bf{F}}}_{k})&={\mathbb{E}}[{{\bf{A}}_k^{\rm {ZF}}}{\hat{\bf{F}}}_{k}{\hat{\bf{F}}}_{k}^H({{\bf{A}}_k^{\rm {ZF}}})^H]\nonumber\\
&\overset{\mathrm{(b)}}{=}{\mathbb{E}}[{{\bf{A}}_k^{\rm {ZF}}}{\bf{I}}_{N_k-N_{\rm J}}({{\bf{A}}_k^{\rm {ZF}}})^H]
={\mbox{Cov}}({{\bf{A}}_k^{\rm {ZF}}})\overset{\mathrm{(c)}}{=}\frac{\eta_{k}{\bf{I}}_{M_{k}}}{N_k-N_J-M_{k}}.
\\[-1 cm]\nonumber
\end{align}
where (b) follows because ${\hat{\bf{F}}}_{k}$ is estimated using the SVD hence all its rows are unit vector and orthogonal to each other, and (c) follows because the use of the estimated beam-forming matrix reduces the dimension the BS-UE equivalent channel from $N_k\times M_{k}$ into $(N_k-N_J)\times M_{k}$. Therefore, the spectral efficiency for when the beam-forming is estimated perfectly is
\begin{align}
C_{k,m}^{\rm ub}=&\log_{2}\big[1+\frac{{P_{\rm T}(N_k-N_J-M_{k})}}{\eta_{k}\sigma_w^2}\big].\qquad\qquad\qquad\quad
\\[-1 cm]\nonumber
\end{align}

For the spectral efficiency lower bound, considering Eq. (\ref{snr_lb_1}) in the worst case when ${\hat{\bf{F}}}_{k}$ is independent of $({\bf{Z}}_{k}{\bf{x}}_{\rm J}+{\bf{w}})$. In this case, the right hand sides of Eq. (\ref{cov5}) and Eq. (\ref{cov2}) demonstrate the covariance of $({{\bf{A}}_k^{\rm {ZF}}}{\hat{\bf{F}}}_{k})$ and $({\bf{Z}}_{k}{\bf{x}}_{\rm J}+{\bf{w}})$ in Eq. (\ref{snr_lb_1}), respectively. Therefore, the lower bound of the spectral efficiency is
\begin{align}
C_{k,m}^{\rm lb}=&\log_{2}\big[1+\frac{{P_{\rm T}(N_k-N_J-M_{k})}}{\eta_{k}(\sigma_w^2+\sum_{j=1}^{N_J}{\eta_{k,j}}\sigma_{{\rm J}_{j}}^2)}\big].\qquad\qquad\qquad\quad
\\[-1 cm]\nonumber
\end{align}
\vspace{-0.5cm}
\appendices
\bibliographystyle{IEEEtran}
\bibliography{IEEEabrv,jamming} 
\end{document}